\documentstyle[12pt]{article}

\begin{document}
\def\thebibliography#1{\section*{REFERENCES\markboth
 {REFERENCES}{REFERENCES}}\list
 {[\arabic{enumi}]}{\settowidth\labelwidth{[#1]}\leftmargin\labelwidth
 \advance\leftmargin\labelsep
 \usecounter{enumi}}
 \def\newblock{\hskip .11em plus .33em minus -.07em}
 \sloppy
 \sfcode`\.=1000\relax}
\let\endthebibliography=\endlist

\hoffset = -1truecm
\voffset = -2truecm


\title{\large\bf
Markov-Yukawa Transversality Principle And 3D-4D Interlinkage Of 
Bethe-Salpeter Amplitudes  
}
\author{
{\normalsize\bf
A.N.Mitra \thanks{e.mail : i) ganmitra@nde.vsnl.net.in ii) anmitra@physics.du.ac.in}
}\\
\normalsize 244 Tagore Park, Delhi-110009, India.}

\date{27 September 2000}

\maketitle

\begin{abstract}
This article is designed to focus attention on the Markov-Yukawa 
Transversality Principle (MYTP) as a novel paradigm for 
an exact 3D-4D interlinkage between the corresponding BSE amplitudes, 
with a closely parallel treatment of $q{\bar q}$ and $qqq$ systems,
stemming from a common 4-fermion Lagrangian mediated by gluon (vector)-like
exchange. This unique feature of MYTP owes its origin to a Lorentz-
covariant 3D support to the BS kernel, which acts as a sort of `gauge
principle' and distinguishes it from most other 3D approaches to strong 
interaction dynamics. Some of the principal approaches in the latter 
category are briefly reviewed so as to set the (less familiar) MYTP  in 
their context. Two specific types of MYTP which provide 3D support to the BSE 
kernel, are considered: a) Covariant Instantaneity Ansatz (CIA); b) Covariant 
LF/NP ansatz (Cov.LF). Both lead to formaly identical 3D BSE reductions 
but produce sharply different 4D structures: Under CIA, the 4D loop 
integrals suffer from Lorentz mismatch of the vertex functions, leading to 
ill-defined time-like momentum integrals, but Cov LF is free from this 
disease. This is illustrated by the pion form factor under Cov LF. The
reconstruction of the 4D $qqq$ wave function is achieved by Green's 
function techniques.  

\end{abstract}

\tableofcontents                                                    

\newpage


\section{A Review Of Strong Interaction Dynamics} 

        Ever since the success of the Tomonaga-Schwinger-Feynman-Dyson
formalism in QED [1], corresponding field-theoretic formulations have been 
in the forefront of strong interaction dynamics since the early fifties,    
the main strategy being to device various `closed' form approaches which
are represented as appropriate  `integral equations'. One of the earliest
efforts in this direction was the Tamm-Dancoff formalism [2] which showed
a great intuitive appeal. In this method, the state vector of the system 
under consideration is Fock-expanded in terms of a complete set of eigen-
functions of the free field Hamiltonian, which was first systematically 
applied by Dyson (+ Cornell collaborators) in the early fifties, to the 
meson-nucleon scattering problem, for a dynamical understanding of the 
`Delta' and other low-energy resonances [3]. It leads to 3D integral 
equations connecting amplitudes for successively higher numbers of meson 
(and/or nucleon-pair) quanta, much as the familiar 4D Schwinger-Dyson 
equations of QED connect (via Ward identities) vertex amplitudes of 
successively higher orders [4]. 

\subsection{3D Reduction of BSE: Quasipotentials, Etc}   

	The 3D Tamm-Dancoff equation (TDE) [3] and the 4D Schwinger-Dyson 
equation (SDE) [4] have been the source of much wisdom underlying the 
formulation of many approaches to strong interaction dynamics. To these one 
should add the Bethe-Salpeter equation (BSE) [5], which is an approximation 
to SDE for the dynamics of a 4D two-particle amplitude, characterized by an
effective (gluon-exchange-like) pairwise interaction, on the lines of a 
"Bethe Second Principle" of the Fifties for the effective N-N interaction, 
but now adapted to the quark level. Although not a fundamental (first
principle) approach, it has attracted more attention in the contemporary 
literature (as the 4D counterpart of the Schroedinger equation) than 
many other approaches.   
\par
	A major bottleneck for the BSE approach has been its resistance to 
a probability interpretation, since the logical demands of its 4D content 
are incompatible with its approximate nature. This has led to many attempts 
at 3D reductions [6-9]: Instantaneous approximation [6]; Quasi-potential 
approaches [7,8]; variants of on-shellness of the associated propagators 
[9]. In [7,9], the starting BSE is 4D in all details, {\it 
{including its kernel}}, but the associated propagators are manipulated in 
various ways to reduce the 4D BSE to a 3D form as a fresh starting point of 
the dynamics; in [8], the old-fashioned 3D perturbation theory is 
reformulated covariantly to give a 3D quasipotential equation. These methods 
are briefly sketched in Sect 2.  
\par 
	At the 4D level, the BSE [5] and its SDE counterpart are still 
the most widely used form of 2-particle dynamics [10], despite the 
problems of probabilistic interpretation. Such equations have been widely 
employed [10] as prototypes of strong interaction dynamics, addressing 
issues of gauge and chiral symmetries, as well as dynamical breaking of 
chiral symmetry ($DB{\chi}S$) via an NJL-type mechanism [11]. In such 
full-fledged field-theoretic approaches, the NJL- mechanism of contact 
interaction logically gives way to space-time extended interactions, and 
the $(DB\chi S)$ corresponds to the use of SDE for the self-energy 
operator [12]. As a general remark, while for conceptual issues of 
formulational self-consistency, there is little alternative to 
the full-fledged 4D equations, their applications to physical systems must
recognize some ground realities. For example, the mass spectra of hadrons 
(which are revealed in Nature as $O(3)$-like [13]), suggest that the role
of the time-like dimension (although an integral part of the dynamics) is   
not on the same footing as that of the space-like dimensions, so that a 
naive expection of $O(4)$-like spectra [14] may be quite misleading. Indeed
this issue is quite central to the very theme of this article, viz., 3D-4D 
interlinkage of BS-amplitudes, and will claim attention throughout.        
\par 
	An alternative form of 2-particle dynamics (which also contributes 
to reducing the effective degrees of freedom from 4D to 3D) is that of Dirac 
constrained Hamiltonian formalism [15], developed by Komar and others [16]. 
The logic of this approach is that constraints $H_i$ have a twin role, viz., 
they not only `constrain the motion' in phase space, but also generate it 
in their `Hamiltonian' capacity. These `constraints' must be mutually 
compatible in the sense $[H_i,H_j]=0$. Such compatibility relations restrict 
the dependence of the interaction on the relative time $t$, and require a 
`reciprocity relation' between the  constituent potentials, something akin 
to Newton's III Law. Such descriptions are valid for both two-boson  
and two-fermion systems [16], in the sense of coupled K-G and Dirac 
equations respectively [17]. (This approach will not be pursued further).   

\subsection{Light Front(LF)/ Null Plane(NP) Dynamics} 

	A powerful form of 3D dynamics came into prominence after 
Weinberg's discovery that the dynamics of the infinite momentum frame [18] 
serves as a cure for many ills in the theory of current algebras, by greatly 
simplifying the rules of calculation of Feynman diagrams of old fashioned 
perturbation theory. In the present context of strong interaction dynamics, 
the great virtue of Weinberg's infinite momentum method [18] lies in the 
simplicity and transparency of the integral equations for multiparticle 
potential scattering problems [19]. Indeed, the structure of the 3-momenta 
$(p_\perp, p_\parallel)$ appearing in this formalism is but a paraphrase for
the standard null-plane variables first introduced by Dirac to project his 
theme [20] that a relativistically invariant Hamiltonian theory can be
based on 3 different classes of initial surfaces (space-like, time-like, 
and null-like). The structure of such a Hamiltonian theory is strongly 
dependent on these respective surface forms  whose "stability groups" (i.e., 
those generators of the Poincare group that leave the initial surface 
{\it invariant}),are $6,6,7$ respectively, thus giving the `highest score' 
$(7)$ to the null-plane dynamics $(x_0=x_3)$ whose `kinematical' generators 
form a closed algebra, and include among others the quantity $P_+=P_0+P_3$ 
(which plays the role of the `mass' term $\eta$ in the Weinberg notation 
[18]). On the other hand, the dual generator $P_-=P_0-P_3$ is the 
`Hamiltonian' of the theory. 
\par
	Leutwyler and Stern [21] gave a covariant 3D formulation of the Dirac 
theme [20] in terms of null-plane variables. A more explicit covariant 
formmulation in the null-plane language was given by Karmanov [22a] using 
diagrammatic techniques with on-shell propagators and spurions, on the
lines of the Kadychevsky approach [8], which has been recently reviewed by 
Carbonell et al [22b],(referred to as KK). All these methods, including the 
Wilson group's [23], give rise to 3D integral equations for a (strongly 
interacting) two-body system, bearing strong resemblance to the other 3D BSE 
forms [6-9] above. Again, there is no getting back to the original 4D BSE 
form,  the nearest connection being a one-way reduction from 4D BSE to 3D on 
the covariant null-plane [22b].

\subsection{Markov-Yukawa Transversality: 3D-4D Interlinkage}    	    

\setcounter{equation}{0}
\renewcommand{\theequation}{1.\arabic{equation}}

Finally we come to a rather novel approach of more recent origin [24-25],  
based on the Markov-Yukawa Transversality Principle (MYTP) [26]. To motivate 
this approach, it is necessary to go back in time to Yukawa's non-local field 
theory [26b] according to which the field variable is a function of both 
$x$ and $p$  (unlike in local field theory where $p$ is absent). Although 
unacceptable for an elementary particle/field, the non-local field theory 
is ideally suited to a $composite$ particle, whose extended structure 
effectively provides for a momentum dependence in the direction of the 
total 4-momentum $P_\mu$. Indeed the Yukawa theory [26b]  was in a way 
the forerunner of a later theory of bi-local fields ${\cal M}(x,y)$ [27] 
for the formulation of the Effective Action for a 2-body dynamical 
system [27]. This approach was employed by the Prevushin group [25] in 
their formulation of the relativistic Coulomb problem in the Salpeter 
approximation [6b] in a covariant form, with the choice of the preferred 
direction governed by the 4-momentum $P_\mu$ of the composite as the 
canonical  conjugate to its c.m. position $X=(x+y)/2$: $P = -i\partial_X $. 
More specifically the MYTP [26] is expressed by the condition [25]:
\begin{equation}
z_\mu {{\partial} \over {\partial X_\mu}} {\cal M}(z,X)=0;\quad z=x-y 
\end{equation}
where the direction $P_\mu$ guarantees an irreducible representation of the 
Poincare' group for the bilocal field ${\cal M}$ [26c]. This condition 
may be viewed as a sort of {\it gauge principle} [26c] which demands 
``redundance'' of the ``relative time'' variable for the bilocal field
${\cal M}(x,P)$ w.r.t. the translation of the relative coordinate:
$$ x_\mu \rightarrow x_\mu + \xi P_\mu $$
which induces the ``gauge trnsformation''
$$ {\cal M}(x_\mu,P_\mu) \rightarrow {\cal M}(x_\mu+ \xi P_\mu, P_\mu) $$
The ``gauge invariance'' now demands that this transformation leave
the bilocal field invariant, which is precisely the condition (1.1) above.    
\par
	The condition (1.1) may also be viewed as equivalent to a covariant 
3D support to the input 4-quark Lagrangian, whence follow the SDE and BSE 
as equations of motion with 3D support to the effective BSE kernel under 
covariant instaneity. More simply, the 3D support ansatz may be postulated 
at the outset for the pairwise BSE kernel $K$ [24] by demanding that 
it be a function of only ${\hat q}_\mu = q-q.PP_\mu/P^2$, which implies that 
${\hat q}.P \equiv 0$. In this approach, the propagators are left untouched 
in their full 4D forms.  This is somewhat complementary to the approaches 
[6-9] (propagators manipulated but kernel left untouched), so that the 
resulting equations [24-25] look rather unfamiliar vis-a-vis 3D BSE's [6-9], 
but it has the advantage of allowing a {\it {simultaneous}} use of both 3D 
and 4D BSE forms via their interlinkage. Indeed what distinguishes the 
MYTP-motivated [26] Covariant Instantaneity Ansatz (CIA) [24] from the 
more familiar 3D reductions of the BSE [6-9] is its capacity for a 
2-way linkage: an $exact$ 3D BSE reduction, and an equally $exact$ 
$reconstruction$ of the original 4D BSE form without extra charge [24]: 
the former to access the observed O(3)-like spectra [13], and the latter 
to give transition amplitudes as 4D quark loop integrals [24]. 
(In the approach of the Pervushin Group [25], however, the built-in  
3D-4D interconnection which follows from MYTP [26], apparently 
remained unnoticed in their final equation). In contrast the more 
familiar methods of 3D [6-9], including the covariant LF [22-23], give 
at most a one-way connection, viz., a $4D\rightarrow 3D$ reduction, 
but not vice versa.  
\par
	At this point it is perhaps worth noting that even the Salpeter 
equation [6b] has (in principle) the ingredients for a reconstruction of the 
4D BS amplitude $\Psi$ in terms of 3D ingredients, insofar as the `instant' 
form, (see eq.(2.1) in Sect 2), of the interaction kernel  is  employed 
on the RHS of the 4D BSE form for the 4D BS amplitude $\Psi$, and duly 
eliminated in favour of the 3D BS amplitude $\psi$, exactly as in CIA [24]. 
This is indeed tantamount to the use of the Transversality Principle [26] 
(albeit non-covariantly), but the possibility of reconstruction of the
4D BS amplitude had apparently remained unnoticed by subsequent workers 
who continued to employ the Salpeter equation [6b] in its 3D form only.   

\subsection{QCD-motivated Effective Lagrangian}

        The Transversality Principle (MYTP) [26] underlying the 3D-4D 
interconnection [24], termed 3D-4D-BSE in the following, of course needs 
supplementing by physical ingredients to govern the structure of the BSE 
kernel, much as a Hamiltonian needs a properly defined `potential'. However 
its canvas is broad enough to accommodate a wide variety of kernels which 
must in turn be governed by independent physical principles. In this respect, 
short of a full-fledged QCD Lagrangian approach, the orthodox view (which we 
adopt) is to stick to an effective 4-fermion Lagrangian as a starting point 
of the dynamics, from which the successive equations of motion (SDE, BSE, 
etc) follow in the standard manner. In this regard, a basic proximity to 
QCD is ensured through a vector-type interaction [12], which while 
maintaining the correct one-gluon-exchange structure in the perturbative 
region, may be fine-tuned to give any desired structure to the mediating 
gluon propagator in the infrared domain as well. Although empirical, it 
captures a good deal of physics in the non-perturbative domain while 
retaining a broad QCD orientation, albeit short of a full-fledged
QCD formulation. More importantly, the non-trivial solution of the SDE 
corresponding to this generalized gluon propagator [12] gives rise to a 
dynamical mass function $m(p)$ [12] as a result of $DB{\chi}S$, w.r.t. an 
input Lagrangian whose chiral invariance stems from a vector-type 4-fermion 
interaction between almost massless $u-d$ quarks. These considerations form
the standard basis for a Lagrangian-based BSE-SDE framework [10] for 
Dynamical Breaking of Chiral Symmetry ($DB{\chi}S$) [11], for a space-time 
extended 4-quark Lagrangian mediated by vector exchange [12]. This generates 
a mass-function $m(p)$ via SDE, which accounts for the 
bulk of the constituent mass of $ud$ quarks. The same BSE-SDE formalism 
[12,10] can be simply adapted [28] to the MYTP [26]-based 3D-4D-BSE formalism
[24] which reproduces 3D spectra of both hadron types [29] under a common 
parametrization [28] for the gluon propagator, with a self-consistent SDE 
determination [28] of the constituent mass; see Sect 3.
\par 
        A BSE-SDE formulation [10] on QCD lines represents a 4D field-
theoretic generalization of `potential models'[30], wherein the generalized 
4-fermion kernel [12] represents the non-perturbative gluon propagator, which 
can be easily adapted [28] to MYTP [26]). The 4D feature of BSE-SDE gives this 
framework a ready access to high energy amplitudes, while its `potential' 
features give it a natural access to hadronic spectra [13]. It has
thus an interpolating role between (low energy) quarkonia models [30], and
(high energy) QCD-SR [31] techniques whose domains are largely complementary;
details may be found in a recent review [32]. 

\subsection{MYTP: Cov Instantaneity vs Cov LF/NP}
 
        While MYTP [26] ensures 3D-4D interconnection [24] under  covariant 
instantaneity ansatz (CIA) in the composite's rest frame [24], its 
applicability is limited by the ill-defined nature of 4D loop integrals which 
acquire time-like momentum components in the exponential/gaussian factors 
associated with the different vertex functions,  due to a `Lorentz-mismatch' 
among the rest-frames of the participating hadronic composites. This problem
is especially serious for triangle loops and above, such as the pion form 
factor, while 2-quark loops [33] just escape this pathology. This problem 
was not explicitly encountered in the light-front (LF/NP) ansatz [34] in an 
earlier study of 4D triangle loop integrals, but this  approach was 
criticized [35] on grounds of non-covariance. The CIA approach [24] which 
made use of MYTP [26], was an attempt to rectify the Lorentz covariance 
defect, but the presence of time-like components in the gaussian factors 
inside triangle loop integrals, e.g., in the pion form factor [36], impeded 
further progress. 
\par
        In an attempt to remedy this situation, a generalization of MYTP
[26] was proposed recently [37] to ensure formal covariance without having 
to encounter time-like components in the gaussian wave functions appearing
inside the 4D loop integrals. The desired generalization was achieved by 
extending the Transversality Principle [26] from the covariant rest frame of 
the (hadron) composite [24], to a {\it {covariantly defined}} light-front [37]
(Cov LF). It was found that while  preserving the 3D-4D BSE interconnection, 
the resulting 3D equation under Cov LF [37] turns out to be formally identical 
to  the old-fashioned null-plane formalism [34,38], so that the latter enjoys 
{\it {ipso facto}} covariance (despite its `looks'). This `covariant' LF/NP 
method [37] compares well with other covariant LF approaches [22-23].     

\subsection{Scope of the Article}

	This article has a 3-fold objective: A) a bird's eye view of some
principal 3D vs 4D dynamical methods for the strong interaction problem that
have been proposed over the last half century; B) Putting in perspective a 
novel property of the Markov-Yukawa Transversality Principle (MYTP), viz.,
a 2-way 3D-4D interconnection in the BS dynamics of 2- and 3-quark hadrons; 
C) Stressing a close parallelism between $q{\bar q}$ and $qqq$ BSE's which 
stem from a common 4-fermion Lagrangian mediated by gluon (vector)-like 
exchange.  Especially noteworthy is the capacity of MYTP [26] to achieve a 
3D-3D interlinkage, a property which has remained obscured from view in 
the contemporary literature, vis-a-vis more familiar approaches to BSE 
and allied forms of dynamics [6-10, 18-23]  which are either 3D or 4D in 
content, but have no provision for any interlinkage between these two 
dimensions.  Our effort will be to stress the practical uses of MYTP in 
the strong interaction regime by virtue of its simplicity in evaluating 
4D loop integrals with arbitrary vertex functions, while providing easy 
access to the spectroscopy sector. To that end, an outline of the dynamical 
structure of some principal 3D methods, [7-9], [18-23], is provided in Sect.2 
as a background for comparison on 4D loop integral techniques, while  on the 
issue of hadron spectra, which are basically $O(3)$-like [13], a comparison 
between non-MYTP [7-9] and MYTP [24-26] forms of dynamics does not bring 
out new physics beyond calibration of the respective parameters.         
\par 
	For a better understanding of the working of MYTP, it will be 
necessary to present two types in parallel for comparison, viz., Covariant 
Instantaneity or CIA [24-25], and Covariant Light-front (Cov LF) [37], which 
demand that the BSE kernel $K$ for pairwise interaction be a function of 
relative momentum $q$ {\it transverse} to the composite 4-momentum in the 
first case [24], and to the Covariant Null-plane in the second [37]. It 
will be shown that both types lead to identical 3D BSE reductions (so that 
their spectral predictions are formally the same), but their reconstructed 
4D vertex functions reveal profound differences in structure: The Lorentz 
mismatch of individual wave functions that characterizes the CIA form [24], 
leading to complex amplitudes [34], disappears in the alternative Cov LF 
approach [37], but in general such integrals are dependent on the 
light-front orientation $n_\mu$, as in other covariant approaches [22-23]. 
To eliminate such terms, a simple prescription of `Lorentz completion' 
seems to suffice to produce an explicitly Lorentz invariant quantity such 
as was shown for the pion form factor [37]; (alternative prescriptions 
exist in other LF/NP formulations [22b]). For a historical perspective,
it is useful to recall that in the old-fashioned NPA approach too [38],  
a very similar result had been found for various types of triangle loop
amplitudes [36], despite a lack of manifest covariance [35] in that approach,
but now MYTP [26] on the covariant null-plane [37] fills this formal gap.
     
\subsection{Outline of Contents}
                    
	Sect.2 briefly outlines some historical approaches to an effective
3D form of strong interaction dynamics: Levy-Salpeter [6]; Logunov-
Tavkhelidze [7a]; Blankenbecler-Sugar [9];  Todorov [7d] ; Weinberg [18]; 
Feynman et al [39]. Sect.3 provides the theoretical framework with a short 
derivation of the BSE-SDE from an input chirally invariant Lagrangian, 
incorporating the original CIA form [25,24] of MYTP [26], on the lines of 
ref.[25] in terms of bilocal fields [27]. It also includes a derivation [28] 
of the dynamical mass function $m(p)$ for an understanding of the constituent 
mass via Politzer additivity [40]. Sect.4  collects some basic results on the 
null-plane formalism due to Leutwyler-Stern [21], especially the role of the 
`Angular Condition' in ensuring a formal $O(3)$-like invariance. From Sect.5 
onwards, the focus is on MYTP [26]-orientation for bringing out its unique 
property which distinguishes it from most other approaches, viz., the
{\it {3D-4D interlinkage}} of BS amplitudes. 
\par 
	Sect.5 gives a comparative view of the working of MYTP on the BSE 
forms in CIA [24] versus Cov LF [37], and outlines the derivation of the 
3D BSE, as well as an explicit reconstruction of the 4D BS wave function in 
terms of 3D ingredients, with 3-momentum ${\hat q} = (q_\perp, q_3)$, where 
the third component emerges as a $P$-dependent one, suitably adapted to the 
CIA [24] or Cov LF [37] respectively. Sec.6 gives a corresponding derivation
for the 3D-4D interconnection for a $qqq$ BSE structure under CIA conditions.  
Sec.7 illustrates, through the calculation of triangle-loop integrals, the 
relative advantage of Cov LF [37] over the CIA [24] version of MYTP [26], in 
producing a well-defined structure for the pion e.m. form factor in 
a fully gauge invariant manner, and illustrating in the process the method 
of `Lorentz-completion' for explicit Lorentz invariance, with the expected 
$k^{-2}$ behaviour at high $k^2$. MYTP also gives a more general structure 
of triangle loop integrals for three-hadron form factors. Sect.8 summarises
our conclusions.  
\par
	     
\section{Quasipotentials And Other 3D Equations} 

The reduction of the 4D BSE for an $N-N$ pair to the 3D level in the 
Instantaneous Approximation was first investigated in the non-adiabatic 
domain of pseudoscalar meson theory (effect of pair-creations included) by 
Levy [6a], who showed that this 3D BSE form is entirely equivalent to the 
corresponding Tamm-Dancoff equations [2] in the same (non-adiabatic) limit. 

\setcounter{equation}{0}
\renewcommand{\theequation}{2.\arabic{equation}}

On the other hand, Salpeter [6b] employed the adiabatic approximation (no 
pair creation effects) to give a systematic 3D reduction of the fermionic 
BSE, using projection operators for large and small components. The adiabatic 
approximation amounts to replacing the propagator $\Delta_F(x-x')$ for the 
exchanged meson by
\begin{equation}\label{2.1}
\Delta_F(x-x') \Rightarrow \delta(x_0-x_0') \int_{-\inf}^{\inf} 
\Delta_F({\bf x-x'}, x_0-x_0') d(x_0-x_0')
\end{equation}  
and simply gives the Yukawa potential between two particles. Similarly, in 
the Instantaneous (adiabatic) Approximation, IA for short,  the 4D wave 
function $\Psi(x)$ = $\Psi({\bf x},t)$ for relative motion of two particles 
becomes simply $\Psi({\bf x}, 0)$. In the momentum representation, these 
statements read respectively as
\begin{equation}\label{2.2}
\Delta_F(k) \Rightarrow \Delta_F({\bf k}) ; \quad 
\psi({\bf q}) = \int dq_0 \Psi({\bf q}, q_0)
\end{equation} 
The Salpeter 3D BSE in the IA for a relativistic hydrogen-atom is [6b]:
\begin{equation}\label{2.3}
(E-H_1({\bf q})-H_2({\bf q})) \chi({\bf q}) = \int d^3k {e^2 \over 
{2\pi^2 ({\bf k-q})^2}} [\Lambda_{1+}\Lambda_{2+} - \Lambda_{1-}\Lambda_{2-}]
\chi({\bf k})
\end{equation}
where the 3D wave function $\chi({\bf q})$ is related to the corresponding 
4D quantity by an equation of the form (2.2), and the symbols $\Lambda_\pm$
are energy projection operators for the large/small components, etc. 
   
\subsection{Logunov-Tavkhelidze Quasipotentials}

A different form of 3D reduction of the 4D BSE was proposed by Logunov-
Tavkhelidze [7a] in the language of Green's functions (G-fns) for 2-particle
scattering whose momentum representation my be written as $G(p_1p_2;p_1'p_2')$
(with indicated 4-momenta before and after), which satisfies a 4D BSE [7a]:
\begin{equation}\label{2.4}
(2\pi)^8 \Delta(p_1)\Delta(p_2) G(p_1p_2;p_1'p_2')= 
{\delta(p_1-p_1')\delta(p_2-p_2')}  \\   \nonumber
+ \int dp_1''dp_2''K(p_1p_2;p_1''p_2'') G(p_1''p_2'';p_1'p_2')    
\end{equation}
where $\Delta(p_i)= p_i^2+m_i^2$, etc. Expressing this equation in c.m. 
($P$) and relative ($q$) 4-momenta, and taking out the $\delta$-fns due 
to the c.m. motion, this equation simplifies to
\begin{equation}\label{2.5}
(2\pi)^4 \Delta(p_1)\Delta(p_2) G(q,q';P) = \delta(q-q') +
\int dq''K(q,q'') G(q'',q;P)
\end{equation}
Next, they defined the 3D G-fn for the relative motion as a double integral 
w.r.t. the two time-like momenta:
\begin{equation}\label{2.6}
{\hat G}({\bf q},{\bf q'};P) = \int q_0 \int q_0' G(q,q';P)
\end{equation}  
Now in operator notation, the 4D BSE (2.5) may be written as $G=G_0+G_0KG$, 
from which the kernel $K$
has the formal representation $K$= $G_0^{-1}-G^{-1}$. The L-T trick [7a]
now consists in using the double integrals on the time-like momenta as in 
eq.(2.6) to formally define the 3D quasipotential ${\hat K}$ as 
\begin{equation}\label{2.7}
{\hat K} \equiv {\hat {G_0}^{-1}} - {\hat G^{-1}}
\end{equation} 
which can be expanded perturbatively in the symbolic form [7a]
\begin{equation}\label{2.8}
{\hat K} = {\hat G_0^{-1}}{\hat {G_0KG_0}}{\hat G_0^{-1}} - 
{\hat G_0^{-1}}{\hat {G_0KG_0KG_0}}{\hat G_0^{-1}} -  ....
\end{equation}
to any desired order of accuracy; [note that the inverse G-fns are just the
self-energy operators]. If $V({\hat q}, {\hat q}';E)$ is the quasi-
potential to a given order of accuracy, then, the BSE satisfied by the 3D 
BS wave function $\psi({\hat q})$ is of the form [7a]:
\begin{equation}\label{2.9}
(E^2 - {\hat q}^2 - m^2) \psi({\hat q}) = \int d^3{\hat q}' 
V({\hat q},{\hat q}';E) \psi({\hat q}')
\end{equation}     
where the `denominator function on the LHS arises from integrating $G_0$=
$\Delta(p_1)^{-1}\Delta(p_2)^{-1}$ w.r.t. $q_0$ and rearranging. 

\subsubsection{Narrow resonances in charged particle systems}

Within the last decade, the L-T theory [7a] has witnessed some interesting
applications [41] to the understanding of `new' narrow $e^+e^-$ resosances 
observed in heavy ion collisions [42]. To that end, the authors [41] have 
employed an equation of the form (2.9) which reads for this system as [41]:
\begin{equation}\label{2.10}
 2\omega(M-2\omega)\phi({\bf p}) = {(2me)^2 \over (2\pi)^3} \int 
{{d^3{\bf p}' \phi({\bf p}')} \over {2\omega'q(M-\omega-\omega'-q+i0)}}
\end{equation}
where $\omega=\sqrt{m^2+{\bf p}^2}$, and $q=\mid {{\bf p}-{\bf p}'} \mid$.   
The results indicate a possible interpretation of the observed peaks [42]
as new quasi-stationary levels arising from the solution of the quasi-
potential equation. More interestingly, they also suggest a close 
relationship of the observed states [42] with the von Neumann-Wigner [43]
levels embedded in the continuum. 

\subsection{Blankenbecler-Sugar Equation}
         
Another type of quasipotential was proposed by Blankenbecler-Sugar [9], 
as follows. The 2-particle scattering amplitude $T(q,q')$ due to a 4D 
potential $V(q,q')$ in the ladder approximation satisfies the BSE [9]:
\begin{equation}\label{2.11}  
T(q,q')= -i(2\pi)^{-4} \int  d^4q''V(q,q'') [m^2+(P/2+q'')^2]^{-1}
[m^2+(P/2-q'')^2]^{-1} T(q'',q')
\end{equation}
where the 2-particle `free' G-fn is exhibited as the product of the two
propagators inside the integral on the RHS. To express this equation in 3D
form, the B-S [9] trick consists first in putting $q'$ on the energy shell, 
which means that $q'_0=0$. and $q'^2=s/4-m^2$, where $s=-P^2$ is the square 
of the c.m. energy. Next, the on-shell part $E_2$ of the free 2-particle 
G-fn is obtained by taking only the $\delta$-fn parts of the two propagators 
which gives rise only to two-particle cuts in the physical region 
\begin{equation}\label{2.12}
E_2(q'') = 2\pi \int ds'(s'-s)^{-1} \delta[m^2+(P'/2+q'')^2]
\delta[m^2+(P'/2-q'')^2]
\end{equation}
where $s'=-P'^2$, and $P'$ has only a fourth component. This works out as 
\begin{equation}\label{2.13}
E_2(q'') = {1 \over 2}\pi\delta(q_0'') [{\sqrt {(q''^2+m^2)}}(q''^2-q^2)]{-1}
\end{equation} 
The balance $R_2$ of the free G-fn is not singular along the positive
cut of the $s$-variable. If it is neglected in the first approximation, and 
only $E_2$ from (2.13) is substituted in (2.11), then after a trivial 
integration over $q_0''$, the resultant 3D equation has the form
\begin{equation}\label{2.14}
T(q,q')=V(q,q') + {1 \over 4} \int {d^3q'' \over {(2\pi)^3}}
{{V(q,q'')T(q'',q')} \over {{\sqrt {m^2+q''^2}} (q''^2-q'^2)}}
\end{equation}    
A comparison of (2.9) and (2.14) shows that although both equations are 
formally 3D in looks, there is a vast difference in their contents: The
L-T [7a] form (2.9) involves only 3-momenta ${\hat q} \equiv {\bf q}$, 
since the Hilbert space has been `truncated' by integrating out over their 
fourth components. The B-S [9] form (2.14) on the other hand has 4-momenta 
formally throughout (no truncation of Hilbert space), except that they are 
on their mass shells ! Thus formal covariance is violated in both equations, 
although in different ways.      

\subsection{Kadychevsky-Todorov Equation}

Still another form of 3D (Lippmann-Schwinger) equation was given by 
Todorov [7d], following the Covariant method of Kadychevsky [8]. In the
Todorov approach [7d], the potential $V_w$ is defined as an infinite power 
series in the coupling constant which fits the perturbative expansion of the 
scattering amplitude $T_w$ for two particles of masses $m_1,m_2$ and 
4-momenta $p_1,p_2$ and $q_1,q_2$ before and after respectively. The 
quantity  $T_w$ in the off-shell regime satisfies the L-S equation [7d]
\begin{equation}\label{2.15}
T_w({\bf p},{\bf q}) +V_w ({\bf p},{\bf q}) + {1 \over {\pi^2 w}}
\int d^3 {\bf k} {{V_w({\bf p},{\bf q})  T_w ({\bf p},{\bf q})} \over 
{{\bf k}^2-b^2-i\epsilon}}  
\end{equation}
where the 3D quantities in the c.m. frame are defined as 
\begin{equation}\label{2.16}
{\bf p}_1 =-{\bf p}_2 ={\bf p}; \quad {\bf q}_1 =-{\bf q}_2 ={\bf q}
\end{equation}
and on the energy shell, the corresponding time-like quantities are
\begin{equation}\label{2.17}
p_{10}+p_{20} = w = q_{10} + q_{20}; \quad -p^2=-q^2=w^2; \quad
4w^2 b^2(w) =\lambda(w^2,m_1^2,m_2^2)
\end{equation}
This equation too has strong resemblance to the L-T equation [7a]. The
corresponding equation for the bound state wave function $\phi({\bf p})$ is
\begin{equation}\label{2.18}
\pi^2 w ({\bf k}^2-b^2(w))\phi({\bf p})= -\int d^3{\bf k} 
V({\bf p},{\bf k}) \phi({\bf k})             
\end{equation}
\par
	Both B-S [9] and Todorov[7d] equations have been extensively 
employed in the literature. 

\subsection{Infinite-Momentum Frame: Weinberg Equation} 

Weinberg [18] observed some remarkable simplifications that occur when the
results of old-fashioned perturbation theory are expressed in a reference
frame in which the total 3-momentum ${\bf P}$ is very large. In this limit, 
the 3-momentum ${\bf p}_n$ of the $n$-th particle may be projected parallel
and perpendicular to ${\bf P}$, and the results collected as follows:
\begin{equation}\label{2.19}
{\bf p}_n=\eta_n{\bf P}+{\bf q}_n; \quad {\bf q}_n.{\bf P}=0; \quad
\sum_n \eta_n = 1; \quad \sum_n {\bf q}_n =0.
\end{equation} 
The quantity $\eta_n > 0$ in this theory, plays the role of `mass' of the
$n$-th particle (in a 3D Schroedinger-type equation), and in the 
$P \rightarrow \infty$ limit, the rules of calculation become very simple: 
all old-fashioned perturbative diagrams passing through negative energy 
intermediate states vanish, while for the contributing diagrams, the 
propagator for an intermediate state $c$ in a transition from $a$ to $b$, 
has the form $2[s_a-s_c+i\epsilon]^{-1}$, where $s$ for any state is the 
usual total c.m. energy squared:
\begin{equation}\label{2.20}
s=\sum_n [{\bf q}_n^2 +m_n^2]/\eta_n; \quad s_a=s_b=s_c, etc.
\end{equation}
Momentum conservation at each vertex is 3D in content:   
\begin{equation}\label{2.21}
(2\pi)^3 \delta(\Delta \sum \eta) \delta^2( \Delta \sum{\bf q})
\end{equation}
in accordance with the conservation of $\eta$ and ${\bf q}$, eq.(2.15).
The Weinberg counterpart of the L-T [7a], B-S [9] and Todorov [7d] equations
(2.9), (2.14) and (2.18) respectively, is the integral equation [18]
\begin{equation}\label{2.22}
<{\bf q}'\eta'\mid T \mid {\bf q}\eta> = 
<{\bf q}'\eta'\mid V \mid {\bf q}\eta>
+ \int d^2q''\int d\eta''{{<{\bf q}'\eta'\mid V \mid {\bf q}\eta>
<{\bf q}'\eta'\mid V \mid {\bf q}\eta>} \over 
{2 (2\pi)^3 [s\eta''(1-\eta'')- {\bf q}''^2- m^2 +i\epsilon]}}
\end{equation}
Although this equation is effectively 3D, and has considerable similarity 
to the corresponding equations of [7a,7d,9] above, there is a big difference,
viz., the angular momentum is no longer a well-defined concept in this 3D
description. This gap was bridged later by Leutwyler-Stern [21] by invoking
the `angular condition' [21], after it became clear that the Weinberg
approach is equivalent to Dirac's [20] null-plane dynamics; see Sect.3 below.  

\subsection{The FKR Model For 2- And 3-Quark Dynamics}

Before ending this Section, we draw attention for historical reasons, to
a unique paper by Feynman and collaborators [39], FKR for short, which gave 
an integrated view of 2- and 3-quark hadron dynamics, and  played a big
role in shaping the direction of strong interaction physics to come. The
importance of the FKR  approach stems, among other things, from the fact 
that these authors were the first to show the way to a unified treatment of 
both 2- and 3-quark hadrons within  a common dynamical framework, which was 
to serve as a model for the future. This paper effectively incorporated all
the relevant aspects of quark dynamics that had been generated piecemeal in
the Sixties, and had by and large come to be accepted, viz., the group 
structure $SU(6) \times O(3)$, the symmetrical quark model, and harmonic 
oscillator classification of hadron states (based on their linear $M^2$-$N$
plots) on the one hand [44], and the mechanism of single-quark transitions,
quark recoil effects, etc, on the other [45]. 
\par
	The FKR model, which made essential use of harmonic confinement, 
sought to give a relativistic meaning to the internal motion of light quarks 
through the following definitions of 2- and 3-quark Hamiltonians [39, 38]:
\begin{equation}\label{2.23}
-K_M=2(p_1^2+p_2^2) + {1 \over 16} \Omega^2 (x_1-x_2)^2 + Const 
\equiv P^2+M_M^2;
\end{equation}
\begin{equation}\label{2.24}
-K_B=3({p_1}^2+{p_2}^2+{p_3}^2) + {1 \over 36} \Omega^2 {\sum_{123} 
{(x_1-x_2)^2}}+ Const \quad \equiv (P^2+M_B^2)
\end{equation}
where $x_{1\mu}$=$i {\partial_{1\mu}}$; $p_1^2 = p_{1\mu}p_{1\mu}$, etc.
The quantity $\Omega$ which is postulated to be the {\it same} for {\it both}
systems, has the significance of the universal Regge slope ($ \approx 1GeV^2$)
as observed [44] in their respective spectra; [Note the geometrical factors
as cofficients in front of the respective kinetic and potential terms above].
The operators $K_{M,B}^{-1}$ are the `free' propagators (albeit with h.o.
confinement) for the mesons and baryons, whose `poles' correspond to the 
eigenvalues (spectra) of their squared masses. The presence of a 
perturbation ${\delta K}$ can be simulated in a standard gauge-invariant 
manner, by the substitutions $p \rightarrow p_\mu -eV_\mu$ or 
$p_\mu \rightarrow p_\mu-g\gamma_5 A_\mu$ for vector and axial vector 
couplings respectively, after rewriting $p_\mu^2$ as 
$(\gamma.p)^2$, while the $i \rightarrow j$  transition amplitudes are 
just $ <h_j\mid {\delta K}\mid h_i>$, by standard rules of quantum mechanics.  
\par
	A major achievement of the FKR model was its success in giving two
distinct types of unification, viz., a common framework for Spectroscopy 
and transition amplitudes; and ii) a unified dynamical treatment  of 
$q{\bar q}$ and $qqq$ hadrons. Both these features represented landmarks in
a dynamical understanding of the quark model, yet the FKR  model failed 
on another count: the `wrong' sign of the time-like momenta in the gaussian 
wave functions for the hadrons was a disease which pointed to an asymmetric 
role of time-like (1D) momenta vis-a-vis the space-like (3D) ones. Attempts 
to cure this disease by a Euclidean treatment (via Wick rotation) [46a]  
failed on the spectroscopy front [13] which reveal only $O(3)$-like spectra, 
while other non-covariant treatments [46b] were not very successful either.  
Nevertheless the lessons from the FKR  model were significant pointers to 
the need to treat the 1D time-like and 3D space-like d.o.f.'s on different 
footings in a future quest for a covariant theory [24-26].  
             
\section{Self-Energy And Vertex Fns Under MYTP}
\setcounter{equation}{0}
\renewcommand{\theequation}{3.\arabic{equation}}

As a first step towards introducing the MYTP [26] theme, we collect in this 
Section some essential machinery for the interconnection between self-energy 
and vertex functions via Schwinger-Dyson (SDE) and Bethe-Salpeter (BSE) 
equations, starting from a chirally invariant Lagrangian characterized by 
a vector-type interaction [12] as a prototype for a gluon-exchange propagator
in the non-perturbative QCD regime [28].  To that end, we shall first outline 
the method of bilocal fields [27] to derive the equations of motion (SDE and 
BSE) from such a Lagrangian, following the Pervushin Group's [25] bilocal
field method, under MYTP [26] conditions of covariant instantaneity [24]. 
This will be followed by a general result connecting self-energy and pion-
quark vertex functions in the chiral limit, i.e., when the pion mass 
vanishes. This result in turn paves the way to a derivation [28], under MYTP 
[26] conditions of Covariant Instantaneity [24], of the mass function $m(p)$ 
whose low momentum limit is the main contributor to the constituent mass, 
via Politzer additivity [40].  

\subsection{Method of Bilocal Fields for BSE-SDE}

The effective action for a system of two interacting massless fermions 
constrained by MYTP [26] is given by [25]
\begin{equation}\label{3.1}
W_{eff}[\psi,{\bar \psi}] = \int d^4x[{\bar \psi}(x)(i\gamma\dot \partial
-m_0)\psi(x) \\ 
+{1 \over 2} \int d^4y (\psi(y){\bar \psi}(x)) K(z^\perp,X)
(\psi(x){\bar \psi}(y))]
\end{equation} 
where  $z=x-y$; $X=(x+y)/2$. $z^\perp$ is the component of $z$ transverse to 
the $P$-direction. Now redefine the action (3.1) in terms of bilocal fields  
${\cal M}$ via the Legendre transformation [27] on the second term to give 
\begin{equation}\label{3.2}
-{1 \over 2} \int d^4xd^4y {\cal M}(x,y) K^{-1}(z^\perp,X){\cal M}(x,y)
+\int d^4xd^4y(\psi(x){\bar \psi}(y) {\cal M}(x,y)
\end{equation}
Then in an obvious short-hand notation [25b], (3.1) may be written as
\begin{equation}\label{3.3}
W_{eff}[{\cal M}]= (\psi{\bar \psi},(-G_0^{-1}+{\cal M}) 
-{1 \over 2} ({\cal M}, K^{-1} {\cal M})
\end{equation}
where $G_0$ is the inverse Dirac operator for the free fermion field. After
quantization over $N_c$ fermion fields and normal ordering, the action takes
the form [25b]
\begin{equation}\label{3.4}
W_{eff}[{\cal M}] = - {1 \over 2}N_c ({\cal M}, K^{-1} {\cal M}) 
+iN_c \sum_{n=1}^{\inf} {1 \over n} \Phi^n
\end{equation}
where $\Phi= G_0{\cal M}$ is a matrix in $(x,y)$ space, and its successive 
powers are defined in the standard matrix fashion. Now for the quantization
of the action (3.4), its minimum is given by
\begin{equation}\label{3.5}
N_c^{-1} {{\delta W_Q({\cal M})} \over {\delta {\cal M}}}
\equiv  -K^{-1} {\cal M} + {1 \over {G_)^{-1}-{\cal M}}} = 0.
\end{equation}
The corresponding `classical' (lowest order) solution for the bilocal field 
is $\Sigma(x-y)$ which depends only on the difference $x-y$ due to the
translation invariance of the vacuum solutions. Next expand the action (3.4)
around the point of minimum ${\cal M}=\Sigma+{\cal M}'$, and denote the 
small fluctuations ${\cal M}'$ as a sum over the complete set of `classical'
solutions $\Gamma$. Then in the next order of extremum, we have:
\begin{equation}\label{3.6}
{{\delta^2{W_Q(\Sigma+{\cal M}')}} \over {\delta {\cal M}^2}} \mid_{{\cal M}'
=0} \dot \Gamma =0
\end{equation}
Eqs.(3.5-6) give respectively the SDE for $\Sigma$ and BSE for $\Gamma$:
\begin{eqnarray}\label{3.7}
\Sigma(x-y) &=& m_0 \delta^4(x-y) + iK(z^\perp,X) G_\Sigma(x-y); \\  \nonumber 
\Gamma      &=& iK(z^\perp,X) \int d^4z_1d^4z_2 G_\Sigma(x-z_1) 
\Gamma(z_1,z_2) G_\Sigma(z_2-y)
\end{eqnarray}
which describe the spectrum of the fermions and composites respectively. In
momentum space these equations for the mass operator and vertex function are  
\begin{equation}\label{3.8}
\Sigma({\hat p}) =m_0 +i \int {d^4q \over {(2\pi)^4}} V({\hat p}-{\hat q})
\gamma\dot \eta G_\Sigma(q) \gamma\dot \eta; \quad \eta_\mu = P_\mu/{|P|}
\end{equation}
\begin{equation}\label{3.9}
\Gamma({\hat k})=i\int {d^4q \over {(2\pi)^4}} V({\hat k}-{\hat q})\gamma\dot
\eta [G_\Sigma(q+P/2)\Gamma(q^\perp)G_\Sigma(q-P/2)]\gamma\dot \eta
\end{equation}
where $G_\Sigma(q)$=$(\gamma\dot q-\Sigma(q^\perp))^{-1}$; $V$ is the scalar 
part of the kernel $K$ with 3D support; ${\hat k}$ is the transverse part
of $k$ w.r.t. the direction $\eta_\mu$ of the total 4-momentum $P_\mu$.
     
\subsection{Self-Energy vs Vertex Fn in Chiral Limit}
 
The formal equivalence of the mass-gap equation (3.8) and the BSE (3.9) 
for a pseudoscalar meson in the chiral limit [12] is now  demonstrated by 
adapting them to a non-perturbative gluon exchange propagator [28] with  
an arbitrary confining form $D(k)$ (not just the perturbative form $k^{-2}$).
The SDE, eq.(3.8), after replacing the color factor $\lambda_1.\lambda_2/4$
by its Casimir value $4/3$,  and a relabelling of symbols [28], now reads 
\begin{equation}\label{3.10}
\Sigma(p) = {4 \over 3}i (2\pi)^{-4} \int d^4k D({\hat k}) 
\gamma\dot \eta S_F'(p-k)\gamma\dot \eta;
\end{equation} 
$S_F'$ is the full propagator related to the mass operator $\Sigma(p)$ by
\begin{equation}\label{3.11}
\Sigma(p) +i \gamma.p = S_F^{-1}(p) = A(p^2)[i\gamma.p + m(p^2)]
\end{equation}
thus defining the mass function $m(p^2)$ in the chiral limit $m_c =0$.
In the same way, eq.(3.9) for the vertex function $\Gamma_H$ of a $q{\bar q}$
hadron $(H)$ of 4-momentum $P_\mu$ made up of quark 4-momenta 
$p_{1,2}= P/2 \pm q$ reads as 
\begin{equation}\label{3.12}
\Gamma_H(q,P)= -{4 \over 3}i(2\pi)^{-4} \int d^4q'D({\hat q}-{\hat q}')
\gamma\dot \eta S_F(q'+ P/2) \Gamma_H(q',P) S_F(q'-P/2) \gamma\dot \eta
\end{equation}
The complete equivalence of (3.10) and (3.12) for the pion case in the 
chiral limit $P_\mu \rightarrow 0$ is easily established. Indeed, with 
the self-consistent ansatz $\Gamma_H$ =$\gamma_5 \Gamma(q)$, eq.(3.12)
simplifies to
\begin{equation}\label{3.13}
\Gamma(q)= {4 \over 3}i (2\pi)^{-4} \int d^4k \gamma\dot \eta S_F'(k-q) 
\Gamma(q-k) S_F'(q-k)\gamma\dot \eta D({\hat k})
\end{equation}
where the replacement $q'= q-k$ has been made. Substitution for $S_F'$ from
(3.11) in (3.13) gives
\begin{equation}\label{3.14}
\Gamma(p) = -{4 \over 3}i (2\pi)^{-4} \int d^4k {{D({\hat k}) \Gamma(p-k)} 
\over {A^2(p-k) (m^2((p-k)^2) + (p-k)^2)}}
\end{equation}
where we have relabelled $q \rightarrow p$. On the other hand substituting
for $S_F'$ (3.11) in (3.10) gives for the mass term of $\Sigma(p)$ the result
\begin{equation}\label{3.15}
A(p^2)m(p^2)= -{4 \over 3}i (2\pi)^{-4} \int d^4k {{D({\hat k})A(q')m(q'^2)} 
\over {A^2(q')(m^2(q'^2) + q'^2)}}
\end{equation}
where $q'=p-k$. A comparison of (3.14) and (3.15) shows their equivalence
with the identification $\Gamma(q)=A(q)m(q^2)$, i.e. the identity of the
vertex and mass functions in the chiral limit, provided $A=1$, (which 
corresponds to the Landau gauge; see [32]). Although obtained here in the
context of MYTP [26] this result is independent of this ansatz. A more 
explicit gauge theoretic derivation of the equations for the self-energy
and vertex functions is given in [32].  

\subsection{Dynamical Mass As $DB{\chi}S$ Solution of SDE}

We end this Section with the definition of the `dynamical' mass 
function of the quark in the chiral limit ($M_\pi=0$) of the pion-quark
vertex function $\Gamma({\hat q})$, in the 3D-4D BSE framework [24,28]. 
The logic of this follows from the BSE-SDE formalism outlined above,
eqs.(3.10-15), for the connection between eq.(3.15) for $m(p)$ and
eq.(3.14) for $\Gamma(q)$ in the limit of zero mass of the pseudoscalar. So,
setting $M=0$ in the (unnormalized) $H q{\bar q}$ vertex function $\Gamma_H$ 
this quantity may be identified with the mass function $m({\hat p})$, in the
limit $P_\mu=0$, where $p_\mu$ is the 4-momentum of either quark;
(note the appearance of the `hatted' momentum). The result is [28,32]
\begin{equation}\label{3.16}
m({\hat p}) = [\omega({\hat p}); {\sqrt 2}p.n]
{{m_q^2 + {\hat p}^2} \over {m_q^2}} \phi({\hat p})
\end{equation}
under CIA and Cov LF respectively. The normalization is such that  in the 
low momentum limit, the constituent $ud$ mass $m_q$ is recovered under CIA
[28], while the corresponding `mass' under Cov LF  is $p_+$ [32].    
 
\section{Null-Plane Preliminaries}

The Weinberg infinite momentum method received a major boost through an
understanding of Bjorken scaling [47a] in deep inelastic scattering, as 
well of the Feynman parton picture [47b] in the same process. The similarity
of the $P \rightarrow \inf$ and the null-plane descriptions became clear 
with the demonstration by Susskind [48] of the $U(2)$ structure of the
LF/NP language wherein the role of `mass' is played by the combination
$p_+=p_0+p_3$, and subsequently a more complete formulation  of null-plane
dynamics by Kogut and Soper [49] within the Hamiltonian formalism in the 
context of field theory.  
\par
	In a different direction, efforts were made to extend the Lorentz
contraction ideas to finite momentum frames, designed to bring out the effect
of Lorentz contraction on cluster form factors as a result of motion [50]. 
In the respect, the role of the Breit frame is particularly interesting
since it gives the best possible overlap [50b] between the initial and final
clusters. The Lorentz contraction factors [50] in turn are the key to an
understanding of `dimensional scaling' [51], especially in a `symmetrized'
version [50d] of the Breit frame [50b] which exactly reproduces the correct
`power counting' [51].  And the Weinberg result [18] is duly recovered in 
the $P \rightarrow \inf$ limit, giving rise to the null-plane dynamics; (for 
more details of these results, see [38]). A more complete formulation of
LF/NP dynamics, albeit with a {\it finite} number of degrees of freedom
was given by Leutwyler-Stern [21] which comes closest to the original Dirac 
[20] spirit, and is summarized below. 

\subsection{Leutwyler-Stern Formalism}

\setcounter{equation}{0}
\renewcommand{\theequation}{4.\arabic{equation}}

Leutwyler-Stern [21] employed a Hamiltonian approach for investigating the 
properties of a relativistic 2-body system with a {\it finite} number of
d.o.f.'s, and postulating a general criterion of `covariance' in the form 
of an operator relation among the mass and spin operators of the system. 
Their formalism is based on the maximum `stability group' (in the Dirac [20]
sense) for the null-plane surface $x_0+x_3=0$, which gives rise to the
following seven `kinematical' generators [38]:
\begin{eqnarray}\label{4.1}
{\cal K}  &=& (P_1,P_2,P_+,E_1,E_2,K_3,J_3); \quad 2P_{\pm}=P_0 \pm P_3;\\  \nonumber  
  2E_1    &=& K_1 + J_2, \quad 2E_2 = K_2 - J_1; 
\quad J_i = {1 \over 2} \epsilon_{ijk} M_{jk}; \quad K_i = M_{0i}.
\end{eqnarray}
The matrix elements of ${\cal K}$ form a closed algebra $(r,s=1,2)$:
\begin{eqnarray}\label{4.2}
[K_3,E_r] &=& -iE_r; \quad [K_3,P_+] = -iP_+; 
\quad [J_3,E_r] = +i\epsilon_{rs} E_s;    \\  \nonumber
[J_3,P_r] &=& +i\epsilon_{rs} P_s; \quad [E_r,P_s] = -i\delta_{rs} P_+ 
\end{eqnarray}
On the other hand, there are 3 `Hamiltonians' which can be chosen in one of 
several different ways. To that end, it is necessary to introduce certain
rotation operators $I_i$ defined by $I_i\mid n>=J_i \mid n>$, on a rest
system $\mid n>$, but extended to states $\mid {\bf p},n>$ defined by
\begin{equation}\label{4.3}
\mid {\bf p},n>= Exp[-i\beta_1E_1-i\beta_2E_2-i\beta_3K_3] \mid n>; \quad
\beta_r=p_r/p_+; \quad \beta_3= \ln (2p_+/M)
\end{equation}
such that ${\bf I}$ commutes with the algebra of ${\cal K}$. More explicitly,
\begin{eqnarray}\label{4.4}
I_i \mid{\bf p},n> &=& Exp[-i\beta_1E_1-i\beta_2E_2-i\beta_3K_3] J_i \mid n>;  \\  \nonumber
 [I_i, I_j]        &=& i\epsilon_{ijk} I_k; \quad [J_i, I_i] = 0.
\end{eqnarray}
In particular, 
\begin{equation}\label{4.5}
I_3=J_3+(E_1P_2-E_2P_1)/P_+ = (W_0+W_3)/P_+; \quad 
W_\mu = {1 \over 2} \epsilon_{\mu\nu\alpha\beta} {P^\nu} M^{\alpha\beta}.
\end{equation}
$W_\mu$ is the Pauli-Lubanski operator, and $M I_r=W_r-P_rW_+/P_+$, where
$r=1,2$. One thus arrives at the `dynamical group' $D=(M,I_i)$, or 
$(M^2, MI_i)$, which has the structure of $U(2)$ [48], because of (3.4).
$D$ is really a 3-member group, since $I_3$ already belongs to ${\cal K}$
by virtue of (3.5). For a particular significance of ${\bf I}$, note its
connection with the non-relativisic Galilei-invariant algebra generated by
the momentum ${\bf P}$ and Galilei boosts ${\bf K}$, viz.,
\begin{equation}\label{4.5}
{\bf I} = {\bf J} + m^{-1} {\bf K} \times {\bf P}; \quad (m=mass)
\end{equation}
Now Galilei invariance of a system is equivalent to the condition that the
corresponding dynamical algebras constitute a unitary representation of 
$U(2)$. In the relativistic case, there is a superficial similarity to the
$U(2)$ structure of $D$, but unlike the NR  case, only the component $I_3$
is now `kinematical', by virtue of (3.5), while $I_{1,2}$ are `dynamical',
and do not have explicit representations by themselves. L-S [21] sought to
bridge this gap by imposing the `covariance' requirement in the form of an 
`angular condition' for the operators $I_i$ as follows:
\begin{equation}\label{4.6}
x_1 M I_1 + x_2 M I_2 + x_L I_3; \quad x_L = P_1 x_1 + p_2 x_2 +P_+ x_-
\end{equation}
which can be shown to be valid on the null-plane $x_+= x_0+x_3=0$ [38].
\par
	The L-S formalism [21] provides a compact support to the longitudinal
momentum $z$ of 2-particle system with constituent 4-momenta $p_{1,2}$:
\begin{equation}\label{4.7}
2z P_+  = p_{1+} - p_{2+}; \quad P_+ = p_{1+} + p_{2+}; \quad
-{1 \over 2} \leq x \leq +{1 \over 2}.
\end{equation}
The internal wave function $\phi$ is defined by 
\begin{equation}\label{4.8}
<{\bf p}_1, {\bf p}_2 \mid {\bf P}, \phi> =
2 P_+ \delta ^3({\bf p}_1+{\bf p}_2-{\bf P}) \phi({\bf q}_\perp, x)
\end{equation}     
where $\phi$ is a matrix in helicity space $(h',h'')$, with the norm:
\begin{equation}\label{4.9}
<\phi \mid \phi> = {1 \over 2} \int {{d^2q_\perp dz} \over {{1\over 4}-z^2}}
\times \sum_{h'h''} {\mid \phi_{h'h''}({\bf q}_\perp, z) \mid}^2
\end{equation}
The L-S [21] structure formally allows the introduction of a 3-vector 
${\bf q}$ and angular momentum ${\bf L}$ for the internal motion of a 
composite of mass $M$ with (equal) constituent masses $m_q$ as
\begin{equation}\label{4.10}
{\bf q} = ({\bf q}_\perp, M x); \quad {\bf L} = 
-i{\bf q} \times {\bf \nabla}_q
\end{equation}
with the phase space
\begin{equation}\label{4.11}
{{d^2 q_{\perp} dz} \over {{1 \over 4} - x^2}} = {{4d^3{\bf q}} \over M};
\quad M^2 = 4(m_q^2 + {\bf q}^2).
\end{equation}    	  
With these definitions of ${\bf q}$ and ${\bf L}$, the theory formally 
preserves the concept of $L$-invariance of a $q{\bar q}$ system despite the
apparently asymmetric treatment meted out to the transverse and longitudinal
components of 3-momenta in the NP or $P=\inf$ formalism [18, 25, 48-9]. This
invariance can be traced to angular condition (4.7). Incidentally, the 
historic FKR model [39], despite its other defects, was found by L-S [21] to 
satisfy the angular condition (4.7).  
\par
	An alternative but more pedigogical recipe to achieve the same end 
was given in [38] via the simpler condition $P.q=0$, to be consistently 
imposed between the total $(P_\mu)$ and relative $(q_\mu)$ 4-momenta of a 
$q{\bar q}$ system, as outlined in subsection 4.2 below [38].   
               
\subsection{An Alternative "Angular Condition" $P.q=0$}

For unequal masses $m_{1,2}$ of the (quark) constituents with 4-momenta
$p_{1,2}$, the total $(P)$ and relative $(q)$ 4-momenta are given by the
Wight-Gaerding definitions [52] 
\begin{equation}\label{4.12}
p_{1,2} = {\hat m}_{1,2}P \pm q; \quad 2{\hat m}_{1,2} = 1 \pm 
{{m_1^2-m_2^2} \over M^2}
\end{equation}
where $M= \sqrt{-P^2}$ is the composite mass. The condition $P\dot q=0$ is
satisfied on the mass shells $m_{1,2}^2 + p_{1,2}^2=0$ of the respective
constituents, by virtue of the Wightman-Gaerding definition (4.13). 
\par
	To link the condition $P.q=0$ with the construction of an
effective 3-vector in the null-plane language, so as to preserve the 
invariance of the angular momentum concept, note that this condition 
translates to the relation $q_-= -q_+ M^2/P_+^2$ which expresses the
component $q_-$ in terms of $q_+$, in a frame $P_\perp =0$, since in this
frame, $P_+P_- = M^2$ on the mass shell of the composite. [The collinearity
condition is not a restriction for a two-body system]. This relation 
then allows a definition of the 3-momentum ${\bf q}$ with the components     
$({\bf q}_\perp, q_3)$, with $q_3= M q_+/P_+$, which preserves the meaning of
${\bf L}$ in the sense of L-S [21], together with NP covariance. For any 
other internal 4-vector $A_\mu$ for the composite, a similar 3-vector 
${\bf A}$ may be defined as $({\bf A}_\perp, A_3)$, with $A_3= M A_+/P_+$,
via the condition $A \dot P=0$. Examples of $A_\mu$ are polarization vectors,
Dirac matrices, etc. Using these techniques, null-plane wave functions of the
L-S type [21] have been constructed and applied to hadronic processes via
quark loops [34]. A more formal mathematical basis for this prescription 
comes from MYTP [26] on the covariant null-plane [37]; see Sect.5 below.   
       
\section{3D-4D BSE Under MYTP: Scalar/Fermion Quarks}

        We now come to our objectives (B) and (C), viz., 3D-4D interlinkage
of BS amplitudes brought about by  MYTP [26], and a unified treatment of
$q{\bar q}$ [24,37] and $qqq$ [53] systems under MYTP conditions. The full 
calculational details with 4-fermion couplings via gluonic propagators have 
been collected in a recent review [32]. However, to bring out the basic 
mathematical structures, we shall derive the 3D-4D interconnection with 
spinless quarks for 2- and 3-quark systems in this and the next sections 
respectively. Further, we shall consider two MYTP methods in parallel for 
comparison: i) Covariant Instantaneity Ansatz (CIA) [24-25]; ii) Covariant 
LF/NP Ansatz (Cov LF) [37], to bring out the structural identity of the 
resulting BSE's for a $q{\bar q}$ system. This will be followed by a 
reconstruction of the 4D BS vertex functions for both types [24, 37] 
as basic building blocks for 4D quark loop calculations. 
             
\setcounter{equation}{0}
\renewcommand{\theequation}{5.\arabic{equation}}                  

\subsection{3D-4D BSE Under CIA: Spinless Quarks}
 
For a self-contained presentation, with unequal mass kinematics [24], let
the quark constituents of masses $m_{1,2}$ and 4-momenta $p_{1,2}$ interact 
to produce a composite hadron of mass $M$ and 4-momentum $P_\mu$. The 
relative 4-momentum $q_\mu$ is related to these by
\begin{equation}\label{5.1}
p_{1,2} = {\hat m}_{1,2}P \pm q; \quad P^2=-M^2; \quad 
2{\hat m}_{1,2}= 1 \pm (m_1^2 - m_2^2)/M^2
\end{equation}
These Wightman-Garding definitions [52] of the fractional momenta 
${\hat m}_{1,2}$ ensure that $q.P=0$ on the mass shells $m_i^2+ p_i^2=0$
of the constituents, though not off-shell. Now define ${\hat q}_\mu$ =
$q_\mu - q.PP_\mu/P^2$ as the relative momentum {\it transverse} to the
hadron 4-momentum $P_\mu$ which automatically gives ${\hat q}.P \equiv 0$,
for all values of ${\hat q}_\mu$. If the BSE kernel $K$ for the 2 quarks
is a function of only these transverse relative momenta, viz. 
$K= K({\hat q}, {\hat q}')$, this is called the ``Cov. Inst.Ansatz (CIA)'' 
[24] which accords with  MYTP [26]. For two scalar quarks with 
inverse propagators $\Delta_{1,2}$, this ansatz gives rise to the following
BSE for the wave fn $\Phi(q,P)$ [24]:
\begin{equation}\label{5.2}
i(2\pi)^4 \Delta_1\Delta_2 \Phi(q,P)= \int d^4q'K({\hat q}, {\hat q}')          
\Phi(q',P); \quad  \Delta_{1,2}= m_{1,2}^2 + p_{1,2}^2
\end{equation} 
The quantities $m_{1,2}$ are the `constituent' masses which are strictly
momentum dependent since they contain the mass function $m(p)$ [12,28], but
may be regarded as constant for low energy phenomena: $m(p) \cong m(0)$.
Further, under CIA, $m(p)= m({\hat p})$, a momentum-dependence which is 
governed by the $DB{\chi}S$ condition [28] (see below). 
\par
        To make a 3D reduction of eq.(5.2), define the 3D wave function 
$\phi({\hat q})$ in terms of the longitudinal momentum $M \sigma$ as
\begin{equation}\label{5.3}
\phi({\hat q}) = \int Md\sigma \Phi(q,P); \quad M\sigma = M q.P/P^2
\end{equation}
using which, eq.(5.2) may be recast as
\begin{equation}\label{5.4}
i(2\pi)^4 \Delta_1 \Delta_2 \Phi(q,P)= \int d^3{\hat q}' K({\hat q},{\hat q}')
\phi({\hat q}'); \quad d^4q' \equiv d^3{\hat q}' M d{\sigma}'
\end{equation}
Next, divide out by $\Delta_1\Delta_2$ in (5.4) and use once again (5.3) to
reduce the 4D BSE form (5.4) to the 3D form 
\begin{equation}\label{5.5}
(2\pi)^3 D({\hat q})\phi({\hat q}) = \int d^3 {\hat q}' K({\hat q},{\hat q}')
\phi({\hat q}'); \quad {{2i\pi}\over {D({\hat q})}} \equiv 
\int {{M d\sigma} \over {\Delta_1\Delta_2}}
\end{equation}
Here $D({\hat q})$ is the 3D denominator function associated with the like
wave function $\phi({\hat q})$. The integration over ${d\sigma}$ is carried 
out by noting pole positions of $\Delta_{1,2}$ in the $\sigma$-plane, where
\begin{equation}\label{5.6}
\Delta_{1,2} = {\omega_{1,2}}^2 - M^2 ({\hat m}_{1,2} \pm \sigma)^2; \quad
{\omega_{1,2}}^2 = m_{1,2}^2 + {\hat q}^2
\end{equation}
The pole positions are given for $\Delta_{1,2}=0$ respectively by
\begin{equation}\label{5.7}
M(\sigma + {\hat m}_1) = \pm \omega_1 \mp i\epsilon; \quad
M(\sigma - {\hat m}_2) = \pm \omega_2 \mp i\epsilon
\end{equation}
where the $(\pm)$ indices refer to the lower/upper halves of the $\sigma$-
plane. The final result for $D({\hat q})$ is expressible symmetrically [24]:
\begin{equation}\label{5.8}
D({\hat q}) = M_{\omega} D_0({\hat q}); \quad {{2} \over {M_{\omega}}} =
{{{\hat m}_1} \over {\omega_1}} + {{{\hat m}_2} \over {\omega_2}} 
\end{equation}
\begin{equation}\label{5.9}
{{1} \over {2}} D_0({\hat q}) = {\hat q}^2 - {{\lambda(m_1^2, m_2^2, M^2)}
 \over {4M^2}}; \quad \lambda = M^4 -2M^2(m_1^2 + m_2^2)+(m_1^2-m_2^2)^2
\end{equation}
The crucial thing for MYTP[26] is now to observe the {\it equality} of 
the RHS of eqs (5.4) and (5.5), thus leading to an {\it {exact 
interconnection}} between the 3D and 4D BS wave functions [24]:
\begin{equation}\label{5.10}
\Gamma({\hat q}) \equiv \Delta_1\Delta_2 \Phi(q,P) = 
{{D({\hat q}\phi({\hat q})} \over {2i\pi}} 
\end{equation}
Eq.(5.10) determines the hadron-quark vertex function $\Gamma({\hat q})$ as 
a product $D\phi$ of the 3D denominator and wave functions, satisfying a 
relativistic 3D Schroedinger-like equation (5.5). The  simultaneous
appearance of the 3D form (5.5) and the 4D form (5.4), leading to their
interconnection (5.10), reveals a two-tier character: The 3D form (5.5) 
gives the basis for making contact with the 3D spectra [13], while the 
reconstructed 4D wave (vertex) function (5.10) in terms of 3D ingredients 
$D$ and $\phi$ can be used for 4D quark-loop integrals in the standard 
Feynman fashion. Note that the vertex function $\Gamma=D\phi/(2i\pi)$ has 
a general structure, independent of the details of the input kernel $K$. 
Further, the $D$-function, eq.(5.8), is universal and well-defined off the 
mass shell of either quark. The 3D wave function $\phi$ is admittedly model-
dependent, but together with $D({\hat q})$, it controls the 3D spectra via 
(5.5), thus offering a direct experimental check on its structure. Both 
functions depend on the single 3D Lorentz-covariant quantity ${\hat q}^2$ 
whose most important property is its positive definiteness for time-lke 
hadron momenta ($M^2 >0$).

\subsection{Cov LF/NP for 3D-4D BSE: Fermion Quarks}

As a preliminary to defining a 3D support to the BS kernel on the light-front 
(LF/NP), on the lines of CIA [24], a covariant LF/NP orientation [37] may be 
represented by the 4-vector $n_\mu$, as well as its dual ${\tilde n}_\mu$, 
obeying the normalizations $n^2 = {\tilde n}^2 =0$ and $n.{\tilde n} = 1$.
In the standard NP scheme (in euclidean notation), these quantities 
are $n=(001;-i)/\sqrt{2}$ and ${\tilde n}=(001;i)/\sqrt{2}$, while the two
other perpendicular directions are collectively denoted by the subscript
$\perp$ on the concerned momenta. We shall try to maintain the $n$-dependence 
of various momenta to ensure explicit covariance; and to keep track of
the usual NP notation $p_{\pm} = p_0 \pm p_3$, our covariant notation is 
normalized to the latter as  $p_+ = n.p \sqrt{2}$; 
$p_- = -{\tilde n}.p \sqrt{2}$, while the perpendicular components continue 
to be denoted by $p_{\perp}$ in both notations.
\par 
        In the same notation as for CIA [24],  the 4th component of the 
relative momentum $q={\hat m}_2 p_1-{\hat m}_1 p_2$,  that should be 
eliminated for obtaining a 3D equation, is now proportional 
to $q_n \equiv {\tilde n}.q$, as the NP analogue [37] of $P.qP/P^2$ in  
CIA [24], where $P=p_1+p_2$ is the total 4-momentum of the hadron. However 
the quantity $q - q_n n$ is still only $q_\perp$, since its square is 
$q^2 - 2 n.q{\tilde n}.q$, as befits $q_\perp^2$ (readily checked against the 
`special' NP frame). We still need a third component $p_3$, for which the 
correct definition turns out to be [37] $q_{3\mu} = z P_n n_\mu$, where 
$P_n$=$P.{\tilde n}$ and $z= q.n/P.n$, which checks with ${\hat q}^2$ = 
$q_\perp^2 + z^2M^2$. We now collect the following definitions/results:
\begin{eqnarray}\label{5.11}
 q_\perp &=& q-q_n n; \quad {\hat q}=q_\perp+ x P_n n; \quad x=q.n/P.n;
 \quad P^2 = -M^2; \\ \nonumber
 q_n,P_n &=& {\tilde n}.(q,P);{\hat q}.n = q.n; \quad {\hat q}.{\tilde n} = 0; 
\quad P_\perp.q_\perp = 0;   \\ \nonumber 
     P.q &=& P_n q.n + P.n q_n; \quad P.{\hat q} = P_n q.n; \quad
{\hat q}^2 = q_\perp^2 + M^2 x^2
\end{eqnarray}        
Now in analogy to CIA, the reduced 3D BSE (wave-fn $\phi$) may be derived
from the 4D BSE (5.2) for spinless quarks (wave-fn $\Phi$) when its kernel 
$K$ is {\it decreed} to be independent of the component $q_n$, i.e., 
$K=K({\hat q},{\hat q}')$, with ${\hat q}$ = $(q_\perp,  P_n n)$, 
in accordance with MYTP [26] condition imposed on the null-plane (NP), 
so that $d^4 q$ = $d^2 q_\perp dq_3 dq_n$. Now define a 3D wave-fn 
$\phi({\hat q})$ = $ \int  d{q_n}\Phi(q)$, as the CNPA  counterpart of the
CIA definition (5.3), and use this result on the RHS of (5.2) to give 
\begin{equation}\label{5.12}
i(2\pi)^4 \Phi(q) = {\Delta_1}^{-1} {\Delta_2}^{-1} 
\int d^3{\hat q}' K({\hat q},{\hat q}')\phi({\hat q}')  
\end{equation}
which is formally the same as eq.(5.4) for CIA above. Now integrate both 
sides of eq.(5.12) w.r.t. $dq_n$ to give a 3D BSE in the variable ${\hat q}$:
\begin{equation}\label{5.13}  
(2\pi)^3 D_n({\hat q}) \phi({\hat q}) =  \int d^2{q_\perp}'d{q_3}'  
K({\hat q},{\hat q}') \phi({\hat q}')
\end{equation}
which again corresponds to the CIA eq.(5.5), except that the function 
$D_n(\hat q)$ is now defined by 
\begin{equation}\label{5.14}
\int d{q_n}{\Delta_1}^{-1} {\Delta_2}^{-1} = 2{\pi}i D_n^{-1}({\hat q})
\end{equation}
and may be obtained by standard NP techniques [38] (Chaps 5-7) as follows. 
In the $q_n$ plane, the poles of $\Delta_{1,2}$ lie on opposite sides of 
the real axis, so that only {\it one} pole will contribute at a time. Taking
the $\Delta_2$-pole, which gives 
\begin{equation}\label{5.15}
2q_n = -{\sqrt 2} q_- = {{m_2^2 + (q_\perp-{\hat m}_2P)^2} 
 \over {{\hat m}_2 P.n - q.n}}
\end{equation}
the residue of $\Delta_1$ works out, after a routine simplification, to 
just $2P.q = 2P.n q_n+2P_n q.n$, after using the collinearity condition 
$P_\perp.q_\perp = 0$ from (5.11). And when the value (5.15) of $q_n$ is 
substituted in (5.14), one obtains (with $P_n P.n = -M^2/2$): 
\begin{equation}\label{5.16}
D_n({\hat q}) = 2P.n ({\hat q}^2 -{{\lambda(M^2, m_1^2, m_2^2)} \over {4M^2}});
\quad {\hat q}^2 = q_\perp^2 + M^2 x^2; \quad x = q.n/P.n      
\end{equation}
Now a comparison of (5.12) with (5.13) relates the 4D and 3D wave-fns: 
\begin{equation}\label{5.17}  
2{\pi}i \Phi(q) = D_n({\hat q}){\Delta_1}^{-1}{\Delta_2}^{-1} \phi({\hat q})
\end{equation}
as the Cov LF counterpart of (5.10) which is valid near the bound state pole. 
The BS vertex function now becomes $\Gamma = D_n \times \phi/(2{\pi}i)$. This
result, though dependent on the LF/NP orientation, is nevertheless formally
{\it covariant}, and closely corresponds to the pedagogical result of the
old LF/NP formulation [38], with $D_n \Leftrightarrow D_+$.    
\par
        A 3D equation similar to the covariant eq.(5.13) above, also obtains 
in alternative LF formulations such as in Kadychevsky-Karmanov [22b] (see 
their eq.(3.48)). However the {\it independent} 4-vector ${\tilde n}_\mu$ 
(which has no counterpart in [22b]), makes this a manifestly covariant 
4D formulation without need for explicit Lorentz transformations [22b]. The 
`angular condition' [21] is also trivially satisfied by the  effective 
3-vector ${\hat q}_\mu$ appearing in the 3D BSE (5.13). A more important
contrast from other null-plane approaches is that the inverse process of 
{\it reconstruction} of the 4D hadron-quark vertex, eq.(5.17)), has no 
counterpart in them [22-23], as these are basically 3D oriented.  
{\it {not vice versa}}.                  
\par
	For fermion quarks with gluonic propagators, the MYTP formulation
needs no new principles, except for certain technical details involving 
slight modifications [54] of the BSE structure for easier handling; 
see [32] for detailed steps. The full 4D wave function $\Psi(P,q)$ may
be expressed as a 4x4 matrix [38,32]:  
\begin{equation}\label{5.18}
\Psi(P,q)= S_F(p_1) \Gamma({\hat q}) \gamma_D S_F(-p_2); \quad
\Gamma({\hat q}) = N_H [1; P_n/M] D({\hat q})\phi({\hat q})/{2i\pi}
\end{equation}   
where $\gamma_D$ is a Dirac matrix which equals $\gamma_5$ for a P-meson, 
$i\gamma_\mu$ for a V-meson, $i\gamma_\mu \gamma_5$ for an A-meson, etc.
The factors in square brackets stand for CIA  and Cov LF values respectively. 
$N_H$ represents the hadron normalization.     

\section{The $qqq$ BSE: 3D-4D Interlinkage}

We now come to the aspect of MYTP [26] that governs the inter-relation of
3D and 4D Bethe-Salpeter amplitudes for $qqq$-systems [53], in keeping 
with a perceived `duality' between meson ($q{\bar q}$) and baryon ($qqq$)
systems which necessitates a parallel treatment between them. In this respect
a fairly comprehensive review of baryon dynamics as a 3-body relativistic
system with full permutation symmetries in all relevant degrees of freedom 
[55] has been given recently [32]. These include: A detailed correspondence 
[56] between $qqq$ and  quark-diquark wave functions; Complex HO techniques 
for the $qqq$ problem [57]; fermionic $qqq$ BSE with the same gluon propagator
for pair $qq$ interactions [29] as employed for $q{\bar q}$ systems [28], 
except for reduction by half due to the color factor; and Green's function 
methods for 3D reduction of the 4D BSE form, plus reconstruction of the 4D 
$qqq$ wave function [53], on the lines of the $q{\bar q}$ problem [24]. 
Within the formalistic scope of this Article however, we shall merely dwell 
on the last item, viz., Green's fn techniques [53] for a 3D reduction of
the 4D BSE, plus {\it reconstruction} of the 4D wave function, for a 
$qqq$ system for three identical spinless quarks, keeping in forefront the 
issue of {\it connectedness} [58] in a 3-particle amplitude whose signal is 
the {\it {absence of any $\delta$-function}} in its structure; (for a 
detailed perspective, see [32]).     

\subsection{Two-Quark Green's Function Under CIA}    

\setcounter{equation}{0}
\renewcommand{\theequation}{6.1.\arabic{equation}}

As a warm up to the method of Green's functions (G-fns),  we first derive 
the 3D-4D interconnection for the corresponding G-fns for 2-particle
scattering of two identical spinless particles, before moving on to the 
3-body problem in the next 2 subsections. For simplicity we shall consider
the G-fns near the bound state poles, so that the inhomogeneous terms may
be dropped. In the notation and phase convention of Section 5, the 4D $qq$ 
Green's fn $G(p_1p_2;{p_1}'{p_2}')$ near a {\it bound} state satisfies a 4D 
BSE (no inhomogeneous term): 
\begin{equation}\label{6.1.1}
i(2\pi)^4 G(p_1 p_2;{p_1}'{p_2}') = {\Delta_1}^{-1} {\Delta_2}^{-1} \int
d{p_1}'' d{p_2}'' K(p_1 p_2;{p_1}''{p_2}'') G({p_1}''{p_2}'';{p_1}'{p_2}');    
\end{equation}
where
\begin{equation}\label{6.1.2}
\Delta_1 = {p_1}^2 + {m_q}^2 , 
\end{equation}
and $m_q$ is the mass of each quark. Now using the relative 4- momentum 
$q = (p_1-p_2)/2$ and total 4-momentum $P = p_1 + p_2$ 
(similarly for the other sets), and removing a $\delta$-function
for overall 4-momentum conservation, from each of the $G$- and $K$- 
functions, eq.(6.1.1) reduces to the simpler form    
\begin{equation}\label{6.1.3}
i(2\pi)^4 G(q.q') = {\Delta_1}^{-1} {\Delta_2}^{-1}  \int d{\hat q}'' 
Md{\sigma}'' K({\hat q},{\hat q''}) G(q'',q')
\end{equation}
where ${\hat q}_{\mu} = q_{\mu} - {\sigma} P_{\mu}$, with 
$\sigma = (q.P)/P^2$, is effectively 3D in content (being orthogonal to
$P_{\mu}$). Here we have incorporated the ansatz of a 3D support for the
kernel $K$ (independent of $\sigma$ and ${\sigma}'$), and broken up the 
4D measure $dq''$ arising from (6.1.1) into the product 
$d{\hat q}''Md{\sigma}''$ of a 3D and a 1D measure respectively. We have 
also suppressed the 4-momentum $P_{\mu}$ label, with $(P^2 = -M^2)$, in 
the notation for $G(q.q')$.
\par

        Now define the fully 3D Green's function 
${\hat G}({\hat q},{\hat q}')$ as [53] 
\begin{equation}\label{6.1.4}
{\hat G}({\hat q},{\hat q}') = \int \int M^2 d{\sigma}d{\sigma}'G(q,q')
\end{equation}
and two (hybrid) 3D-4D Green's functions ${\tilde G}({\hat q},q')$,
${\tilde G}(q,{\hat q}')$ as
\begin{equation}\label{6.1.5}
{\tilde G}({\hat q},q') = \int Md{\sigma} G(q,q'); \quad
{\tilde G}(q,{\hat q}') = \int Md{\sigma}' G(q,q');
\end{equation} 
Next, use (6.1.5) in (6.1.3) to give    
\begin{equation}\label{6.1.6}
i(2\pi)^4 {\tilde G}(q,{\hat q}') = {\Delta_1}^{-1} {\Delta_2}^{-1} 
\int dq'' K({\hat q},{\hat q}''){\tilde G}(q'',{\hat q}')  
\end{equation}
Now integrate both sides of (6.1.3) w.r.t. $Md{\sigma}$ and use the result 
\begin{equation}\label{6.1.7}
\int Md{\sigma}{\Delta_1}^{-1} {\Delta_2}^{-1} = 2{\pi}i D^{-1}({\hat q});
\quad D({\hat q}) = 4{\hat \omega}({\hat \omega}^2 - M^2/4);\quad 
{\hat \omega}^2 = {m_q}^2 + {\hat q}^2 
\end{equation}
to give a 3D BSE w.r.t. the variable ${\hat q}$, while keeping the other 
variable $q'$ in a 4D form:
\begin{equation}\label{6.1.8}  
(2\pi)^3 {\tilde G}({\hat q},q') = D^{-1} \int d{\hat q}''  
K({\hat q},{\hat q}'') {\tilde G}({\hat q}'',q')
\end{equation}
A comparison of (6.1.3) with (6.1.8) gives the desired connection between 
the full 4D $G$-function and the hybrid ${\tilde G({\hat q}, q')}$-function: 
\begin{equation}\label{6.1.9}  
2{\pi}i G(q,q') = D({\hat q}){\Delta_1}^{-1}{\Delta_2}^{-1}
{\tilde G}({\hat q},q')
\end{equation}
Again, the symmetry of the left hand side of (6.1.9) w.r.t. $q$ and $q'$ 
allows rewriting the right hand side with the roles of $q$ and $q'$ 
interchanged. This gives the dual form   
\begin{equation}\label{6.1.10}  
2{\pi}i G(q,q') = D({\hat q}'){{\Delta_1}'}^{-1}{{\Delta_2}'}^{-1}
{\tilde G}(q,{\hat q}')
\end{equation}
which on integrating both sides w.r.t. $M d{\sigma}$ gives
\begin{equation}\label{6.1.11}  
2{\pi}i{\tilde G}({\hat q},q') = D({\hat q}'){{\Delta_1}'}^{-1}
{{\Delta_2}'}^{-1}{\hat G}({\hat q},{\hat q}'). 
\end{equation}
Substitution of (6.1.11) in (6.1.9) then gives the symmetrical form
\begin{equation}\label{6.1.12}  
(2{\pi}i)^2 G(q,q') = D({\hat q}){\Delta_1}^{-1}{\Delta_2}^{-1}
{\hat G}({\hat q},{\hat q}')D({\hat q}'){{\Delta_1}'}^{-1}
{{\Delta_2}'}^{-1}
\end{equation}
Finally, integrating both sides of (6.1.8) w.r.t. $M d{\sigma}'$, we 
obtain a fully reduced 3D BSE for the 3D Green's function:
\begin{equation}\label{6.1.13}  
(2\pi)^3 {\hat G}({\hat q},{\hat q}') = D^{-1}({\hat q} \int d{\hat q}''
K({\hat q},{\hat q}'') {\hat G}({\hat q}'',{\hat q}')
\end{equation}
Eq.(6.1.12) which is valid near the bound state pole, expresses the desired 
connection between the 3D and 4D forms of the Green's functions; and 
eq(6.1.13) is the determining equation for the 3D form. A spectral analysis 
can now be made for either of the 3D or 4D Green's functions in the 
standard manner, viz., 
\begin{equation}\label{6.1.14}  
G(q,q') = \sum_n {\Phi}_n(q;P){\Phi}_n^*(q';P)/(P^2 + M^2) 
\end{equation}
where $\Phi$ is the 4D BS wave function. A similar expansion holds for 
the 3D $G$-function ${\hat G}$ in terms of ${\phi}_n({\hat q})$. Substituting
these expansions in (6.1.12), one immediately sees the connection between 
the 3D and 4D wave functions in the form:
\begin{equation}\label{6.1.15}  
2{\pi}i{\Phi}(q,P) = {\Delta_1}^{-1}{\Delta_2}^{-1}D(\hat q){\phi}(\hat q)
\end{equation}
whence the BS vertex function becomes $\Gamma$ = $D \times \phi/(2{\pi}i)$
as found in [24]. We shall make free use of these results, taken as $qq$ 
subsystems, for our study of the $qqq$ $G$-functions in subsects.6.2-3.  

\subsection{3D BSE Reduction for $qqq$ G-fn}
\setcounter{equation}{0}
\renewcommand{\theequation}{6.2.\arabic{equation}}

        As in the two-body case, and in an obvious notation for various 
4-momenta (without the Greek suffixes), we consider the most general 
Green's function $G(p_1 p_2 p_3;{p_1}' {p_2}' {p_3}')$ for 3-quark 
scattering {\it near the bound state pole} (for simplicity) which allows       
us to drop the various inhomogeneous terms from the beginning. Again we 
take out an overall delta function $\delta(p_1 + p_2 + p_3 - P)$ from the
$G$-function  and work with {\it two} internal 4-momenta for each of the 
initial and final states defined as follows [54b]:
\begin{equation}\label{6.2.1}  
{\sqrt 3}{\xi}_3 =p_1 - p_2 \ ; \quad  3{\eta}_3 = - 2p_3 + p_1 +p_2
\end{equation}
\begin{equation}\label{6.2.2}  
P = p_1 + p_2 + p_3 = {p_1}' + {p_2}' + {p_3}'
\end{equation}
and two other sets ${\xi}_1,{\eta}_1$ and ${\xi}_2,{\eta}_2$ defined by 
cyclic permutations from (6.2.1). Further, as we shall consider pairwise
kernels with 3D support, we define the effectively 3D momenta ${\hat p}_i$, 
as well as the three (cyclic) sets of internal momenta 
${\hat \xi}_i,{\hat \eta}_i$, (i = 1,2,3) by [54b]:
\begin{equation}\label{6.2.3}
{\hat p}_i = p_i - {\nu}_i P \ ;\quad  {\hat {\xi}}_i = {\xi}_i - s_i P\  ;
\quad
{\hat {\eta}}_i - t_i P 
\end{equation}
\begin{equation}\label{6.2.4}  
{\nu}_i = (P.p_i)/P^2\  ;\quad s_i = (P.\xi_i)/P^2 \ ;\quad t_i = 
(P.\eta_i)/P^2 \end{equation}
\begin{equation}\label{6.2.5}  
{\sqrt 3} s_3 = \nu_1 - \nu_2 \ ;\quad 3 t_3 = -2 \nu_3 + \nu_1 + \nu_2 \ 
\ ( + {\rm cyclic permutations})
\end{equation}

The space-like momenta ${\hat p}_i$ and the time-like ones $\nu_i$ 
satisfy [54b] 
\begin{equation}\label{6.2.6}  
{\hat p}_1 + {\hat p}_2 + {\hat p}_3 = 0\  ;\quad \nu_1 + \nu_2 + \nu_3 = 1
\end{equation}
Strictly speaking, in the spirit of covariant instantaneity, we should 
have taken the relative 3D momenta ${\hat \xi},{\hat \eta}$ to be in the 
instantaneous frames of the concerned pairs, i.e., w.r.t. the rest frames
of $P_{ij} = p_i +p_j$; however the difference between the rest frames of 
$P$ and $P_{ij}$  is small and calculable [54b], while the use of a common 
3-body rest frame $(P = 0)$ lends considerable simplicity and elegance to 
the formalism.   
\par
        We may now use the foregoing considerations to write down the BSE 
for the 6-point Green's function in terms of relative momenta, on closely 
parallel lines to the 2-body case. To that end note that the 2-body 
relative momenta are $q_{ij} = (p_i - p_j)/2 = {\sqrt 3}{\xi_k}/2$, where 
(ijk) are cyclic permutations of (123). Then for the reduced $qqq$ Green's
function, when the {\it last} interaction was in the (ij) pair, we may use 
the notation $G(\xi_k \eta_k ; {\xi_k}' {\eta_k}')$, together with `hat' 
notations on these 4-momenta when the corresponding time-like components 
are integrated out. Further, since the pair $\xi_k,\eta_k$ is 
{\it {permutation invariant}} as a whole, we may choose to drop the index 
notation from the complete $G$-function to emphasize this symmetry as and 
when needed. The $G$-function for the $qqq$ system satisfies, in the 
neighbourhood of the bound state pole, the following (homogeneous) 4D BSE
for pairwise $qq$ kernels with 3D support:
\begin{equation}\label{6.2.7}  
i(2\pi)^4 G(\xi \eta ;{\xi}' {\eta}') = \sum_{123}
{\Delta_1}^{-1} {\Delta_2}^{-1} \int d{{\hat q}_{12}}'' M d{\sigma_{12}}''
K({\hat q}_{12}, {{\hat q}_{12}}'') G({\xi_3}'' {\eta_3}'';{\xi_3}' {\eta_3}')
\end{equation}
where we have employed a mixed notation ($q_{12}$ versus $\xi_3$) to stress
the two-body nature of the interaction with one spectator at a time, in a 
normalization directly comparable with eq.(6.1.3) for the corresponding 
two-body problem. Note also the connections 
\begin{equation}\label{6.2.8}  
\sigma_{12} = {\sqrt 3}{s_3}/2   ;\quad 
{\hat q}_{12} = {\sqrt 3}{{\hat \xi}_3}/2  ; \quad {\hat \eta}_3 = 
-{\hat p}_3, \quad etc 
\end{equation}  
The next task is to reduce the 4D BSE (6.2.7) to a fully 3D form through a 
sequence of integrations w.r.t. the time-like momenta $s_i,t_i$ applied 
to the different terms on the right hand side, {\it {provided both}} 
variables are simultaneously permuted. We now define the following fully 
3D as well as mixed (hybrid) 3D-4D $G$-functions according as one or more 
of the time-like $\xi,\eta$ variables are integrated out:
\begin{equation}\label{6.2.9}  
{\hat G}({\hat \xi} {\hat \eta};{\hat \xi}' {\hat \eta}') = 
\int \int \int \int ds dt ds' dt' G(\xi \eta ; {\xi}' {\eta}')  
\end{equation}  
which is $S_3$-symmetric.
\begin{equation}\label{6.2.10}  
{\tilde G}_{3\eta}(\xi {\hat \eta};{\xi}' {\hat \eta}') = 
\int \int dt_3 d{t_3}' G(\xi \eta ; {\xi}' {\eta}');
\end{equation}  
\begin{equation}\label{6.2.11}  
{\tilde G}_{3\xi}({\hat \xi}  \eta;{\hat \xi}' {\eta}') = 
\int \int ds_3 d{s_3}' G(\xi \eta ; {\xi}' {\eta}');
\end{equation} 
The last two equations are however {\it not} symmetric w.r.t. the 
permutation group $S_3$, since both the variables ${\xi,\eta}$ are not 
simultaneously transformed; this fact has been indicated in eqs.(8.2.10-11) 
by the suffix ``3" on the corresponding (hybrid) ${\tilde G}$-functions,
to emphasize that the `asymmetry' is w.r.t. the index ``3". We shall term 
such quantities ``$S_3$-indexed", to distinguish them from $S_3$-symmetric 
quantities as in eq.(6.2.9). The full 3D BSE for the ${\hat G}$-function is 
obtained by integrating out both sides of (6.2.7) w.r.t. the $st$-pair 
variables $ds_i d{s_j}' dt_i d{t_j}'$ (giving rise to an $S_3$-symmetric 
quantity), and using (6.2.9) together with (6.2.8) as follows:
\begin{equation}\label{6.2.12}  
(2\pi)^3 {\hat G}({\hat \xi} {\hat \eta} ;{\hat \xi}' {\hat \eta}') = 
\sum_{123} D^{-1}({\hat q}_{12}) \int d{{\hat q}_{12}}'' 
K({\hat q}_{12}, {{\hat q}_{12}}'') {\hat G}({\hat \xi}'' {\hat \eta}'';
{\hat \xi}' {\hat \eta}')  
\end{equation}   
This integral equation for ${\hat G}$ which is the 3-body counterpart of
(6.1.13) for a $qq$ system in the neighbourhood of the bound state pole, 
is the desired 3D BSE for the $qqq$ system in a {\it {fully connected}}
form, i.e., free from delta functions. Now using a spectral decomposition 
for ${\hat G}$ 
\begin{equation}\label{6.2.13}   
{\hat G}({\hat \xi} {\hat \eta};{\hat \xi}' {\hat \eta}')
= \sum_n {\phi}_n( {\hat \xi} {\hat \eta} ;P)
{\phi}_n^*({\hat \xi}' {\hat \eta}';P)/(P^2 + M^2)
\end{equation}   
on both sides of (6.2.12) and equating the residues near a given pole
$P^2 = -M^2$, gives the desired equation for the 3D wave function $\phi$ 
for the bound state in the connected form:
\begin{equation}\label{6.2.14}   
(2\pi)^3 \phi({\hat \xi} {\hat \eta} ;P) = \sum_{123} D^{-1}({\hat q}_{12})
\int d{{\hat q}_{12}}'' K({\hat q}_{12}, {{\hat q}_{12}}'')
\phi({\hat \xi}'' {\hat \eta}'' ;P)
\end{equation}   
Now the $S_3$-symmetry of $\phi$ in the $({\hat \xi}_i, {\hat \eta}_i)$ pair 
is a very useful result for both the solution of (6.2.14) {\it and} for the 
reconstruction of the 4D BS wave function in terms of the 3D wave function 
(6.2.14), as is done in the subsect.6.3 below.

\subsection{Reconstruction of 4D $qqq$ Wave Function}

\setcounter{equation}{0}
\renewcommand{\theequation}{6.3.\arabic{equation}}
\par

        We now attempt to {\it re-express} the 4D $G$-function given by 
(6.2.7) in terms of the 3D ${\hat G}$-function given by (6.2.12), as the 
$qqq$ counterpart of the $qq$ results (6.1.12-13). To that end we adapt 
the result (6.1.12) to the hybrid Green's function  of the (12) subsystem 
given by ${\tilde G}_{3 \eta}$, eq.(6.2.10), in which the 3-momenta 
${\hat \eta}_3,{{\hat \eta}_3}'$ play a parametric role reflecting the 
spectator status of quark $\# 3$, while the {\it active} roles are played 
by $q_{12}, {q_{12}}' = {\sqrt 3}(\xi_3,{\xi_3}')/2$, for which the analysis 
of subsect.6.1 applies directly. This gives 
\begin{equation}\label{6.3.1} 
(2{\pi}i)^2 {\tilde G}_{3 \eta}(\xi_3 {\hat \eta}_3; 
{\xi_3}' {{\hat \eta}_3}') 
= D({\hat q}_{12}){\Delta_1}^{-1}{\Delta_2}^{-1}
{\hat G}({\hat \xi_3} {\hat \eta_3}; {\hat \xi_3}' {\hat \eta_3}')
D({{\hat q}_{12}}'){{\Delta_1}'}^{-1}{{\Delta_2}'}^{-1}
\end{equation}
where on the right hand side, the `hatted' $G$-function has full 
$S_3$-symmetry, although (for purposes of book-keeping) we have not 
shown this fact explicitly by deleting the suffix `3' from its 
arguments. A second relation of this kind may be obtained from (6.2.7)
by noting that the 3 terms on its right hand side may be expressed in 
terms of the hybrid ${\tilde G}_{3 \xi}$ functions vide their definitions 
(6.2.11), together with the 2-body interconnection between $(\xi_3,{\xi_3}')$ 
and $({\hat \xi}_3,{{\hat \xi}_3}')$ expressed once again via (6.3.1), but 
without the `hats' on $\eta_3$ and ${\eta_3}'$. This gives
\begin{eqnarray}\label{6.3.2}
({\sqrt 3} \pi i)^2 G(\xi_3 \eta_3; {\xi_3}'{\eta_3}')
&=& ({\sqrt 3} \pi i)^2 G(\xi \eta; {\xi}'{\eta}')\nonumber\\
&=& \sum_{123} {\Delta_1}^{-1}{\Delta_2}^{-1} (\pi i {\sqrt 3})
\int d{{\hat q}_{12}}'' M d{\sigma_{12}}''
K({\hat q}_{12}, {{\hat q}_{12}}'') 
G({\xi_3}'' {\eta_3}'';{\xi_3}' {\eta_3}')\nonumber\\   
&=& \sum_{123} D({\hat q}_{12}) {\Delta_1}^{-1}{\Delta_2}^{-1}
{\tilde G}_{3 \xi}({\hat \xi}_3  \eta_3; {{\hat \xi}_3}' {{\eta}_3}')
{{\Delta_1}'}^{-1} {{\Delta_2}'}^{-1}  
\end{eqnarray}
where the second form exploits the symmetry between $\xi,\eta$ and 
$\xi',\eta'$. 
\par
        At this stage, unlike the 2-body case, the reconstruction of the
4D Green's function is {\it {not yet}} complete for the 3-body case, as 
eq.(6.3.2) clearly shows. This is due to the {\it truncation} of Hilbert 
space implied in the ansatz of 3D support to the pairwise BSE kernel $K$ 
which, while facilitating a 4D to 3D BSE reduction without extra charge, 
does {\it not} have the {\it complete} information to permit the {\it reverse}
transition (3D to 4D) without additional assumptions. To fill up this gap
in this ``inverse" mathematical problem, we look for a suitable  ansatz for 
${\tilde G}_{3 \xi}$ on the RHS of (6.3.2) in terms of {\it known} quantities,
so that the reconstructed 4D $G$-function satisfies the 3D equation (6.2.12) 
exactly, as a check-point. We therefore seek a structure of the form 
\begin{equation}\label{6.3.3}
{\tilde G}_{3 \xi}({\hat \xi}_3  {\eta}_3; {{\hat \xi}_3}' {{\eta}_3}')
= {\hat G}({{\hat \xi}_3} {\hat \eta}_3; {{\hat \xi}_3}' {{\hat \eta}_3}')
\times F(p_3, {p_3}')    
\end{equation}
where the unknown function $F$ must involve only the momentum of the 
spectator quark $\# 3$. A part of the $\eta_3, {\eta_3}'$ dependence has 
been absorbed in the ${\hat G}$ function on the right, so as to satisfy 
the requirements of $S_3$-symmetry for this 3D quantity [53]. 
\par

        As to the remaining factor $F$, it is necessary to choose its
form in a careful manner so as to conform to the conservation of 
4-momentum for the {\it free} propagation of the spectator between two
neighbouring vertices, consistently with the symmetry between $p_3$ 
and ${p_3}'$. A possible choice consistent with these conditions is:
\begin{equation}\label{6.3.4}
F(p_3, {p_3}') = C_3 {\Delta_3}^{-1} {\delta}(\nu_3 - {\nu_3}') 
\end{equation}
Here ${\Delta_3}^{-1}$ represents the ``free" propagation of quark $\# 3$ 
between successive vertices, while $C_3$ represents some residual effects 
which may at most depend on the 3-momentum ${\hat p}_3$, but must satisfy 
the main constraint that the 3D BSE, (6.2.12), be {\it explicitly} satisfied.
\par
        To check the self-consistency of the ansatz (6.3.4), integrate
both sides of (6.3.2) w.r.t. $ds_3 d{s_3}' dt_3 d{t_3}'$ to recover the 
3D $S_3$-invariant ${\hat G}$-function on the left hand side. Next, in 
the first form on the right hand side, integrate w.r.t. $ds_3 d{s_3}'$ 
on the $G$-function which alone involves these variables. This yields
the quantity ${\tilde G}_{3 \xi}$. At this stage, employ the ansatz 
(6.3.4) to integrate over $dt_3 d{t_3}'$. Consistency with the 3D BSE, 
eq.(6.2.12), now demands 
\begin{equation}\label{6.3.5}
C_3 \int \int d\nu_3 d{\nu_3}' {\Delta_3}^{-1} \delta(\nu_3 - {\nu_3}')
= 1 ; (since dt = d\nu) 
\end{equation}
The 1D integration w.r.t. $d\nu_3$ may be evaluated as a contour 
integral over the propagator ${\Delta}^{-1}$ , which gives the pole 
at $\nu_3 = {\hat \omega}_3/M$, (see below for its definition). Evaluating 
the residue then gives 
\begin{equation}\label{6.3.6}
C_3 = i \pi / (M {\hat \omega}_3 ) ;  \quad
{{\hat \omega}_3}^2 = {m_q}^2 + {{\hat p}_3}^2
\end{equation}
which will reproduce the 3D BSE, eq.(6.2.12), {\it exactly}! Substitution
of (6.3.4) in the second form of (6.3.2) finally gives the desired 3-body 
generalization of (6.1.12) in the form 
\begin{equation}\label{6.3.7}
3 G(\xi \eta; \xi' \eta') = \sum_{123} D({\hat q}_{12}) \Delta_{1F} 
\Delta_{2F} D({{\hat q}_{12}}') {\Delta_{1F}}' {\Delta_{2F}}' 
{\hat G}({\hat \xi_3} {\hat \eta_3}; {\hat \xi_3}' {\hat \eta_3}')
[\Delta_{3F} / (M \pi {\hat \omega}_3)]      
\end{equation}
where for each index, $\Delta_F = - i {\Delta}^{-1}$ is the 
Feynman propagator.  To find the effect of the ansatz (6.3.4) on the 4D 
BS wave function $\Phi(\xi \eta; P)$, we do a spectral reduction 
like (6.2.13) for the 4D Green's function $G$ on the LHS of 
(6.3.2). Equating the residues on both sides gives the desired 4D-3D 
connection between $\Phi$ and $\phi$:
\begin{equation}\label{6.3.8}
\Phi(\xi \eta; P) = \sum_{123} D({\hat q}_{12}){\Delta_1}^{-1}{\Delta_2}^{-1}
\phi ({\hat \xi} {\hat \eta}; P) \times 
\sqrt{{\delta(\nu_3 -{\hat \omega}_3/M)} \over{M {\hat \omega}_3 
{\Delta}_3}} 
\end{equation}
{\it defines} the 4D wave fn in terms of piecewise vertex fns $V_i$, as   
\begin{equation}\label{6.3.9}
\Phi(p_1p_2p_3) \equiv {{V_1+V_2+V_3} \over {\Delta_1 \Delta_2\Delta_3}}    
\end{equation}
From (6.3.8-9), we infer the baryon-$qqq$ vertex function $V_3$ corresponding
to the `last' interaction in the $12$-pair  as 
\begin{equation}\label{6.3.10}
V_3 = D({\hat q}_{12}\phi({\hat \xi},{\hat \eta}) \times {\sqrt 
{2\Delta_3\delta({\nu_3}^2 M^2-{\hat \omega}_3^2)}}
\end{equation}
and so on cyclically. (The argument of the 
$\delta$-function inside the radical for $V_3$ simplifies to $p_3^2+m_q^2$). 
This expression had been obtained earlier from intuitive considerations [54b]. 
\par
        To account for the appearance of the 1D $\delta$-fn under radical
in (6.3.10), it is explained elsewhere [53] that it has nothing to do with 
connectedness [58] as such, but merely reflects a `dimensional mismatch' due 
to the 3D nature of the pairwise kernel $K$ [24] imbedded in a 4D Hilbert 
space. (For a physical explanation, see [53]). A further self-consistency 
check on (6.3.10), is found by taking the limit of a point interaction, 
which amounts to setting $K=Constant$, when the radical (expectedly)
disappears, and gives a Lorentz-invariant result [53], in agreement with 
the so-called NJL-Faddeev (contact) model [59] for 3-particle scattering.      
For the fermion $qqq$ case with pairwise gluonic interactions, the details
may be found in [60], wherein the strength of the `color' $qq$ interaction
[29] is half of that of $q{\bar q}$ [28]. For brevity, we skip the MYTP [26]
derivation of the 4D $qqq$ vertex function under Cov LF [37] conditions, which 
parallels that for the 2-body case [Sect.5], except for the remark that  
the old-fashioned LF/NP treatment [38] gives the same results as the more 
formal Cov LF treatment in Sect.5, so that a similar Cov LF form for $qqq$ 
dynamics should be expected [38], with $D({\hat q}) \rightarrow 
D_n({\hat q})$, etc in (6.3.10). 

\section{Triangle Loops Under MYTP On Cov LF/NP} 

	In this Section, we shall illustrate the MYTP techniques on the 
covariant light-front to bring out the main feature, viz., structure of the 
triangle loop integrals free from the anomalies of time-like momenta in the 
product of gaussian vertex functions, such as complexities in the pion
form factor [36] (see Sects. 1 and 5). To that end, we shall mainly 
consider the mathematical structure of the P-meson form factor, followed  
by a brief stetch of the structure of 3-hadron form factors, in the next
few subsections, leaving routine calculational details to [37,32].       

\newpage

\subsection{Pion Form Factor by Cov LF/NP Method}  

\setcounter{equation}{0}
\renewcommand{\theequation}{7.\arabic{equation}}

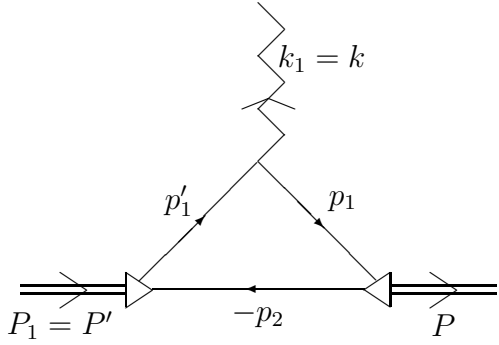
\begin{figure}[h]

\vspace{0.5in}

\begin{picture}(450,50)(-50,100)
\put (110,60){\line(1,0){40}}
\put (110,63){\line(1,0){40}}

\put (250,60){\line(1,0){40}}
\put (250,63){\line(1,0){40}}

\put (125,68){\line(3,-2){10}}
\put (125,54.5){\line(3,2){10}}

\put(125,48){\makebox(0,0){$P_1=P'$}}
\put(270,48){\makebox(0,0){$P$}}
\put(170,95){\makebox(0,0){$p'_1$}}

\put (265,68){\line(3,-2){10}}
\put (265,54.5){\line(3,2){10}}

\put (150,68){\line(3,-2){10}}
\put (150,54.5){\line(3,2){10}}
\put (150,68){\line(0,-1){13}}

\put (240,61.5){\line(3,-2){10}}
\put (240,61.5){\line(3,2){10}}
\put (250,68){\line(0,-1){13}}

\multiput(200,110)(0,20){3}{\line(1,1){10}}
\multiput(210,120)(0,20){3}{\line(-1,1){10}}
\put (194,130){\line(5,2){10}} 
\put (204,134){\line(5,-2){10}}

\put(224,149){\makebox(0,0){$k_1=k$}}

\put(200,52){\makebox(0,0){$-p_2$}}

\put (200,110){\vector(1,-1){25}} 
\put (224.5,85.2){\line(1,-1){20}}

\put (155,64.7){\vector(1,1){25}}
\put (179.5,89.5){\line(1,1){20}}

\put(232,95){\makebox(0,0){$p_1$}}

\put (240,61.5){\vector(-1,0){45}}
\put (215,61.5){\line(-1,0){55}}

\end{picture}
\vspace{1.0in}
\caption{Triangle loop for e.m. vertex}

\end{figure}

\vspace{0.5in}

        Using fig.1 above, and an identical one with $1 \rightarrow 2$, 
(c.f., figs. 1a,1b of [34b]), the Feynman amplitude for the 
$h \rightarrow h'+ \gamma$ transition  is given by [34b]
\begin{equation}\label{7.1}                
2{\bar P}_\mu F(k^2)= 4(2\pi)^4 N_n(P)N_n(P') e{\hat m}_1 \int d^4 T_\mu^{(1)}
{{D_n({\hat q})\phi({\hat q})D_n({\hat q}')\phi({\hat q}')} \over 
{\Delta_1 \Delta_1' \Delta_2}} + [1 \Rightarrow 2];
\end{equation}   
\begin{equation}\label{7.2}
4T_\mu^{(1)} = Tr [\gamma_5 (m_1-i\gamma.p_1) i\gamma_\mu (m_1-i\gamma.p_1')
\gamma_5 (m_2 + i\gamma.p_2)]; \quad \Delta_i = m_i^2 + p_i^2; 
\end{equation}
\begin{equation}\label{7.3}
p_{1,2} = {\hat m}_{1,2}P \pm q; \quad p_{1,2}'={\hat m}_{1,2}P' \pm q'
\quad p_2 = p_2'; \quad P-P'=p_1-p_1'=k; \quad 2{\bar P} = P+P'.
\end{equation}
After evaluating the traces and simplifying, $T_\mu$ becomes
\begin{equation}\label{7.4}
T_\mu^{(1)}= (p_{2\mu}-{\bar P}_\mu)[{\delta m}^2-M^2-\Delta_2] -k^2p_{2\mu}/2
+(\Delta_1-\Delta_1')k_\mu/4
\end{equation} 
The last term in (7.4) is non-gauge invariant, but it does not survive the
integration in (7.1), since the coefficient of $k_\mu$, viz., 
$\Delta_1-\Delta_1'$ is antisymmetric in $p_1$ and $p_1'$, while the rest 
of the integrand in (7.1) is symmetric in these two variables. Next, to bring
out the proportionality of the integral (7.1) to ${\bar P}_\mu$, it is
necessary to resolve $p_2$ into the mutually perpendicular components 
$p_{2\perp}$, $(p_2.k/k^2)k$ and $(p_2.{\bar P}/{\bar P}^2){\bar P}$, of
which the first two will again not survive the integration, the first due
to the angular integration, and the second due to the antisymmetry of
$k=p_1-p_1'$ in $p_1$ and $p_1'$, just as in the last term of (7.4). The third
term is explicitly proportional to ${\bar P}_\mu$, and is of course 
gauge invariant since ${\bar P}.k =0$. (This fact had been anticipated while 
writing the LHS of (7.4)). Now with the help of the results
\begin{equation}\label{7.5}
p_2.{\bar P}= -{\hat m}_2 M^2 -\Delta_1/4 - \Delta_1'/4; \quad
2{\hat m}_2 = 1-(m_1^2-m_2^2)/M^2; \quad {\bar P}^2=-M^2-k^2/4,
\end{equation} 
it is a simple matter to integrate (7.1), on the lines of Sec.5, noting 
that terms proportional to $\Delta_1 \Delta_2$ and $\Delta_1' \Delta_2$
will give zero, while the non-vanishing terms will get contributions 
only from the residues of the $\Delta_2$-pole, eq.(5.15). Before collecting
the various pieces, note that the 3D gaussian wave functions $\phi,\phi'$,
as well as the 3D denominator functions $D_n, D_n'$, do {\it not} depend 
on the time-like components $p_{2n}$, so that no further pole contributions
accrue from these sources. (It is this problem of time-like components of
the internal 4-momenta inside the gaussian $\phi$-functions under the CIA
approach [24], that had plagued a earlier CIA study of triangle diagrams 
[36]). To proceed further, it is now convenient 
to define the quantity ${\bar q}.n = p_2.n -{\hat m}_2 {\bar P}.n$ to 
simplify the $\phi$- and $D_n$- functions. To that end define the symbols: 
\begin{equation}\label{7.6}
(q,q') = {\bar q} \pm {\hat m}_2 k/2; \quad  z_2 = {\bar q}.n/{{\bar P}.n}; 
\quad {\hat k}= k.n/{{\bar P}.n}; \quad (\theta_k, \eta_k)=1 \pm {\hat k}^2/4
\end{equation}
and note the following results of pole integration w.r.t. $p_{2n}$ [38]:
\begin{equation}\label{7.7}
\int dp_{2n}{1 \over {\Delta_2}} [1/{\Delta_1}; 1/{\Delta_1'};
1/(\Delta_1 \Delta_1')] = [1/D_n; 1/D_n'; 2p_2.n/(D_nD_n')]
\end{equation}          
Details of further calculation of the form factor are given in [37].
 An essential result is the normalizer $N_n(P)$ of the hadron, 
obtained by setting $k_\mu=0$, and demanding that $F(0)=1$. The reduced 
(Lorentz-invariant) normalizer $N_H= N_n(P)P.n/M$ is given by [32,37]:   
\begin{equation}\label{7.8}
N_H^{-2}=2M (2\pi)^3 \int d^3{\hat q}
e^{-{\hat q}^2/\beta^2} [(1+{\delta m}^2/M^2)({\hat q}^2-\lambda/{4M^2})
+2{\hat m}_1{\hat m}_2 (M^2-{\delta m}^2)]           
\end{equation}
where the internal momentum ${\hat q}=(q_\perp,Mz_2)$ is formally a 3-vector,
in conformity with the `angular condition' [21]. The corresponding 
expression for the form factor is [32, 37]: 
\begin{equation}\label{7.9}
F(k^2)= 2M N_H^2 (2\pi)^3 exp[-{(M{\hat m}_2{\hat k}/\beta)^2/{4\theta_k}}] 
(\pi \beta^2)^{3/2} {\eta_k \over {\sqrt \theta_k}}{\hat m}_1 G({\hat k}) +
[1 \Rightarrow 2]            
\end{equation}
where $G({\hat k})$ is a function of ${\hat k}$; see eqs.(A.12-13) of [32]. 

\subsection{`Lorentz Completion' for $F(k^2)$}

The expression (7.9) for $F(k^2)$ still depends on the null-plane
orientation $n_\mu$ via the dimensionless quantity ${\hat k}$ =
$k.n/P.n$ which while having simple Lorentz transformation properties, 
is nevertheless {\it not} Lorentz invariant by itself. To make it explicitly 
Lorentz invariant, we shall employ a simple method of `Lorentz completion' 
which is merely an extension of the `collinearity trick' empolyed at the
quark level, viz., $P_\perp.q_\perp$ = 0; see eq.(5.11). Note that this
collinearity ansatz has already become reduntant at the level of the 
Normalizer $N_H$, eq.(7.8), which owes its Lorentz invariance to the 
integrating out of the null-plane dependent quantity $z_2$ in (7.8). This
is of course because $N_H$ depends only on one 4-momentum (that of a
{\it {single hadron}}), so that the collinearity assumption is exactly
valid. However the form factor $F(k^2)$ depends on {\it {two independent}}
4-momenta $P,P'$, for which the collinearity assumption is non-trivial,
since the existence of the perpendicular components cannot be wished away!
Actually the quark-level assumption $P_\perp.q_\perp$ = 0 has, so to say, got 
transferred, via the ${\hat q}$-integration in eq.(7.9), to the {\it {hadron
level}}, as evidenced from the ${\hat k}$-dependence of $F(k^2)$; therefore
an obvious logical inference is to suppose this ${\hat k}$-dependence  
to be the result of the  collinearity ansatz $P_\perp.P_\perp'$ = 0 at the
hadron level. Now, under the collinearity condition, one has
\begin{equation}\label{7.10}
P.P' = P_\perp.P_\perp'+ P.n P'.{\tilde n} + P'.n P.{\tilde n}
 = P.n P_n'+ P'.n P_n ; \quad P.{\tilde n} \equiv P_n.
\end{equation}
Therefore `Lorentz completion'(the opposite of the collinearity ansatz) 
merely amounts to reversing the direction of the above equation by supplying
the (zero term) $P_\perp.P_\perp'$ to a 3-scalar product to render it a 
4-scalar! Indeed the process is quite unique for 3-point functions such as 
the form factor under study, although for more involved cases (e.g., 
4-point functions), further assumptions may be needed. 
\par
        In the present case, the prescription of Lorentz completion is 
relatively simple, being already contained in eq.(7.10). Thus since
$P,P'$= ${\bar P} \pm k/2$, a simple application of (7.10) gives 
\begin{eqnarray}\label{7.11}
k.n k_n     &=& +k^2; \quad {\bar P}.n {\bar P}_n = -M^2 -k^2/4; \\  \nonumber
{\hat k}^2  &=& {{4k^2} \over {4M^2+k^2}}= 4\theta_k-4=4-4\eta_k 
\end{eqnarray}    
This simple prescription for ${\hat k}$ automatically ensures the 4D 
(Lorentz) invariance of $F(k^2)$ at the hadron level. (For comparison with
alternative methods [22b], see [37]).   

\subsection{QED Gauge Corrections to $F(k^2)$} 
\par
        While the `kinematic' gauge invariance of $F(k^2)$ has already been 
ensured in Sec.7.1 above, there are additional contributions to the
triangle loops - figs.1a and 1b of [34b] - obtained by  inserting the photon 
lines at each of the two vertex blobs instead of on the quark lines themselves.
These terms arise from the demands of QED gauge invariance, as pointed out 
by Kisslinger and Li [61] in the context of two-point functions, and are
simulated by inserting exponential phase integrals with the e.m. currents.
However, this method (which works ideally for {\it point} interactions) is 
not amenable to {\it extended} (momentum-dependent) vertex functions, and
an alternative strategy is needed, as described below.     
\par
        The way to an effective QED gauge invariance lies in the simple-minded 
substitution $p_i-e_iA(x_i)$ for each 4-momentum $p_i$ (in a mixed $p,x$ 
representation) occurring in the structure of the vertex function. This 
amounts to replacing each ${\hat q}_\mu$ occurring in $\Gamma({\hat q})$ = 
$D({\hat q})\phi({\hat q})$, by ${\hat q}_\mu -e_q {\hat A}_\mu$, where 
$e_q = {\hat m}_2 e_1 -{\hat m}_1 e_2$, and keeping only first order terms
in $A_\mu$ after due expansion. Now the first order correction to 
${\hat q}^2$ is $-e_q{\hat q}.{\hat A}- e_q{\hat A}.{\hat q}$, which 
simplifies on substitution from  eq.(7.11) to 
\begin{equation}\label{7.12}
-2e_q \tilde q.A \equiv -2e_q A_\mu [{\hat q}_\mu -{\hat q}.n {\tilde n}_\mu 
+ P.{\tilde n}{\hat q}.n n_\mu /P.n]
\end{equation}    
The net result is a first order correction to $\Gamma({\hat q})$ of amount 
$e_q j({\hat q}).A$ where  
\begin{equation}\label{7.13}
j({\hat q})_\mu = -4M_>{\tilde q}_\mu\phi({\hat q}) (1-({\hat q}^2 -
\lambda/{4M^2})/{2\beta^2})
\end{equation}
The contribution to the P-meson form factor from this hadron-quark-photon 
vertex (4-point) now gives the QED gauge correction to the triangle loops, 
in the form of a similar function $F_1(k^2)$ which works out as [37]:
\begin{equation}\label{7.14}
F_1(k^2)= 4(2\pi)^4 N_H^2 e_q {\hat m}_1 M_>^2 \int d^4q(M^2-{\delta m}^2)
\phi\phi'[{{D_n'{\hat q}.P}\over {\Delta_1'\Delta_2 P'.n}}
+ {{D_n{\hat q}'.P}\over {\Delta_1\Delta_2 P.n}}] + \{1 \Rightarrow 2\}
\end{equation}
where $M_> = \sup\{M; m_1+m_2\}$ [37], and the common factor $2{\bar P}_\mu$
has been extracted as for $F(k^2)$ in (7.1). Note that $e_q$ is antisymmetric
in `1' and `2', signifying a change of sign when $\{1 \Rightarrow 2\}$ is
added on the RHS. The pole integration of $F_1(k^2)$ now yields a result  
like (7.9) for $F(k^2)$; see [37] for details.
\par     
	The large and small $k^2$ limits of $F(k^2)$ and $F_1(k^2)$ 
are on expected lines, and we summarise only the final results for 
completeness [37]. For large $k^2$, the functions $F(k^2)$ and $F_1(k^2)$ 
both yield the correct asymptotic form $C/k^2$, where $C= 0.35 GeV^2$, to
be compared with the experimental value [62a] $0.50 \pm 0.10$, and the
(perturbative) QCD value [63] $8\pi \alpha_s f_\pi^2 = 0.296$. 
\par
	For low $k^2$, on the other hand, an expansion of $F,F_1$ in powers
of $k^2$ yields a value of the charge radius $R$ according to $<R^2>$ =
$-{\nabla_k}^2 (F(k^2)+F_1(k^2))$ in the $k^2=0$ limit. Of the two functions,
only $F(k^2)$ contributes in this limit [37]. The numerical values for the
kaon and pion radii, vis-a-vis experiment [62b], are    
\begin{equation}\label{7.15}
R_K = 0.63 fm \quad vs (0.53 fm) ; \quad R_\pi = 0.661 fm \quad vs (0.656 fm). 
\end{equation}

\subsection{Three-Hadron Couplings Via Triangle-Loops}

\begin{figure}[h]
\vspace{3.5in}

\begin{picture}(450,50)(-50,-170)

\thinlines

\put (210,-34){\vector(-3,2){53}}
\put (137,15){\line(3,-2){20}}
\put(162,-16){\makebox(0,0){${\bar t}(-p_2)$}}
\put (137,15){\vector(3,2){50}}
\put(164,46){\makebox(0,0){$t(p_1)$}}
\put (187,48){\line(3,2){22}}

\put (213,68){\line(5,0){50}}
\put (213,64){\line(5,0){50}}
\put (213,-40){\line(5,0){50}}
\put (213,-36){\line(5,0){50}}
\put(253,80){\makebox(0,0){$T(P_1)$}}
\put(253,-52){\makebox(0,0){${\bar T}(P_2)$}}

\put (213,58){\vector(0,-5){58}}
\put(229,15){\makebox(0,0){$q(p_3)$}}
\put (213,0){\line(0,-5){30}}

\thicklines

\put (70,15){\vector(2,0){50}}
\put(85,22){\makebox(0,0){$F(P)$}}
\put (107,15){\line(1,0){30}}

\put (243,66){\line(-3,-1){10}}
\put (243,66){\line(-3,1){10}}

\put (243,-38){\line(-3,-1){10}}
\put (243,-38){\line(-3,1){10}}

\put (213,-28){\line(0,-1){20}}

\put (213,76){\line(0,-1){20}}

\put (213,56){\line(-1,2){5}}
\put (213,76){\line(-1,-2){5}}

\put (213,-28){\line(-1,-2){5}}
\put (213,-48){\line(-1,2){5}}

\end{picture}
\vspace{-1.25in}
\caption{3-hadron coupling}

\end{figure}
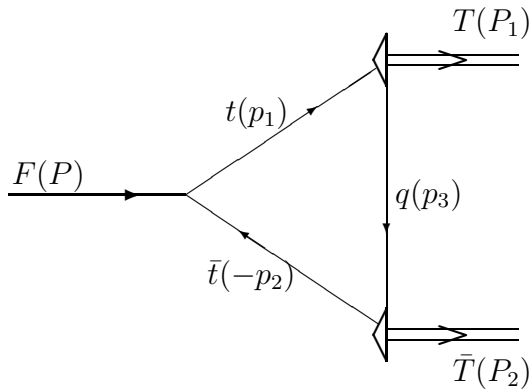

\vspace{0.25in}

For a large class of hadronic processes like $H \rightarrow H'+H''$ and
$H \rightarrow H'+\gamma$, the quark triangle loop [64] represents the 
lowest order ``tree'' diagram for their evaluation. Criss-cross gluonic 
exchanges inside the triangle-loop  are not important for this description 
in which the hadron-quark vertices, as well as the quark propagators are 
{\it {both non-perturbative}}, and thus take up a lion's share of non-
perturbative effects. This is somewhat similar to the `dynamical perturbation 
theory' of Pagels-Stokar [65], in which criss-cross diagrams are neglected. 
\par
        We now indicate in the barest outline, the structure of the 3-hadron 
loop integral for the most general case of unequal mass kinematics 
$m_1 \neq m_2 \neq m_3$, while referring for notational details to ref.[32].
The full structure of the 3-hadron amplitude may be written down from fig.2
above (c.f., fig.1 of [64]), just as for the e.m. form factor (7.1): 
\begin{equation}\label{7.16}
A(3H)= {{2i} \over {\sqrt 3}} (2\pi)^8 \int d^4p_i 
\Pi_{123}{{\Gamma_i({\hat q}_i)} \over {\Delta_i(p_i)}} 
\end{equation}
exhibiting cyclic symmetry, where the normalized vertex function $\Gamma_i$ 
in CNPA [41] is given in an obvious notation by eq.(5.18) as
\begin{equation}\label{7.17}
\Gamma_i({\hat q}_i)= N_i (2\pi)^{-5/2}
D_i({\hat q}_i)\phi_i({\hat q}_i); \quad D_i= 2M_i({\hat q}_i^2
-{{\lambda(M_i^2, m_j^2,m_k^2)} \over {4M_i^2}})
\end{equation}
where the `reduced' denominator function $D_i$ = $D_{i+}M_i/P_{i+}$ and the 
(invariant) normalizer $N_{iH}$ is $N_i$. The color factor and the effect of 
reversing the loop direction are given by $2/{\sqrt 3}$, etc [38,64]. 
the overall BS normalizer [38]. 
\par
        To evaluate (7.16), we first write the cyclically invariant measure:
\begin{equation}\label{7.18}
d^4p_i = d^p_{\perp} {1 \over 2} d(x_i^2) M_i^2 dy_i; \quad 
x_i = p_{i+}/P_{i+}; \quad y_i = p_{i-}/P_{i-}
\end{equation}   
The cyclic invariance of (7.18) ensures that it is enough to take any
index, say $2$, and first do the pole integration w.r.t. the $y_2$ variable
which has a pole at $y_2=\xi_2$$ \equiv $$\omega_{2\perp}^2/(M_2^2 x_2)$.
The process can be repeated, by turn, over all the indices and the results
added. Note that the $\phi$-functions do {\it not} include the time-like
$y_i$ variables under CNPA [37], so that the residues from the poles arise
from only the propagators. The crucial thing to note is that the denominator
functions $D_1$ and $D_3$ sitting at the opposite ends of the $p_2$-line (c.f.
Fig.1 of [64]) will {\it {cancel out}} the residues from the complementary 
(inverse) propagators $\Delta_3$ and $\Delta_1$ respectively. Indeed by
substituting the pole value $y_2=\xi_2$, in $\Delta_{1,3}$, the corresponding
residues in an obvious notation work out as [32]:
\begin{equation}\label{7.19}
\Delta_{1;2}=\xi_2 n_{32}M_2^2 + x_2 n_{23}M_3^2 - 2{\hat \mu}_{21}M_3^2; \\\\       
\Delta_{3;2}=-\xi_2 n_{12}M_2^2 - x_2 n_{21}M_1^2 - 2{\hat \mu}_{23}M_1^2
\end{equation}
It is then found, with a short calculation [32], that
\begin{equation}\label{7.20}
{{D_3({\hat q}_3)} \over {\Delta_{1;2}}} = 2M_3x_2n_{23};\quad
{{D_1({\hat q}_1)} \over {\Delta_{3;2}}} = 2M_1x_2 n_{21} 
\end{equation}
which shows the precise cancellation mechanism between the $D_i$-functions 
and the residues of the propagators $\Delta_i$ at the $\Delta_2$ pole. This
mechanism thus eliminates [24, 64] the (overlapping) Landau-Cutkowsky poles 
that would otherwise have caused free propagation of quarks in the loops. 
The same procedure is then repeated cyclically for the other two terms 
arising from the $\Delta_{3,1}$ poles. Collecting the factors, the result
of all the 3 contributions is compactly expressible as [64, 32]:
\begin{equation}\label{7.21}
A(3H)= 8{\sqrt {{2\pi} \over 3}} \Sigma_{123} \int \int M_2 {n_{23} n_{21}}
\pi^2 dx_2 d\xi_2 x_2^2 [TR]_2 D_{2}({\hat q}_2) {\Pi_{123}{M_iN_i\phi_i}}   
\end{equation}
where the limits of integration for both variables are 
$-\inf<(\xi_2,x_2)<+\inf$, since these are governed, not by the on-shell 
dynamics of standard LF methods [22-23], but by off-shell 3D-4D BSE.  
The difference from [64] (under CIA [24]) arises from using CNPA [37] which
has ensured that the (gaussian) functions $\phi_i$ on the RHS of (7.21) are
now free from time-like momenta (unlike in CIA [24,64]). 
\par
	Eq.(7.21) is the central result of this exercise. Its general nature 
stems from the use of unequal mass kinematics at both the quark and hadron 
levels, which greatly enhances its applicability to a wide class of problems 
involving 3-hadron couplings, either as complete processes by themselves 
(such as in decay processes) or as parts of bigger diagrams in which 3-hadron 
couplings serve as basic building blocks. What makes the formula particularly 
useful for general applications is its explicit Lorentz invariance which has 
been achieved through the simple method of `Lorentz Completion' on the lines 
of sect.7.2 for the e.m. form factor of P-mesons; for more details, see [32]. 
\par
	As regards two- quark loops, such as for $SU(2)$ mass splittings of 
P-mesons [33b], and the mixing of $\rho$ and $\omega$ off-shell propagators 
[33a], the distinction between CIA [24] and CNPA [37] is less sharp, (no
time-like momentum problems in the overlap integrals). The same holds for
one-quark loops, e.g., in the problem of vacuum condensates. For  more
details of these processes, as well as for other references, see [32]. 
            
\section{Retrospect And Conclusions}

To set the relatively unfamiliar MYTP [26] in the context of other
BSE-SDE type approaches to strong interaction dynamics, Sections 1-2   
have attempted a panoramic view of several standard approaches to 3D BSE 
reductions [6-9], as a background for putting in perspective its unique 
feature of $exact$ 3D-4D interconnection of BS amplitudes of both  
$q{\bar q}$ and $qqq$ types in closely parallel fashions. Further
background methodology is provided in Sections 3-4, the former for 
a general derivation of the equations of 
motion in interlinked BSE-SDE form from an input 4-fermion Lagrangian for
`current' quarks, under MYTP conditions [25], and the latter for some
essential Light-front techniques, such as the `angular condition' [21]. 
With this background, Sect.5     
outlines a comparative derivation of the MYTP-controlled 3D-4D interlinkage 
of $q{\bar q}$ Bethe-Salpeter amplitudes under both CIA [24] and Cov LF [37]
conditions. And in keeping with the parallelism between the 2- and 3-quark
treatments, Sect.6 gives a similar derivation for the $qqq$ sustem. Now 
the 3D-4D interlinkage which charactizes MYTP, unlike other approaches  
[6-9,22-23], gives rise to a natural two-tier description [38], the 3D BSE 
form  being appropriate for making contact with the hadron spectra [13], 
while the reconstructed 4D BSE yields a vertex function which allows the use 
of standard Feynman diagrams for 4D loop integrals. The examples of the
pion form factor and more general 3-hadron couplings in Section 7 
illustrate the advantages of Cov-LF [37] over CIA [24] in producing 
well-defined triangle loop integrals, except for a (less serious) 
problem of dependence on the `null-plane orientation' which can be handled 
through a simple device of `Lorentz completion' to yield an explicitly 
Lorentz-invariant structure. 
\par
	In keeping with its mathematical (formalistic) emphasis of
this Article, we have refrained from discussing the phenomenological 
applications, but it has been shown that the canvas of MYTP [26] is broad
enough to accommodate additional physical principles. In particular, the
physical basis chosen for detailed presentation, has been a QCD motivated 
4-fermion Lagrangian (with an effective gluonic propagator) which generates 
the BSE-SDE structure by breaking its chiral symmetry dynamically 
($DB{\chi}S$) [11-12], formulated within an MYTP [26] framework.    
\par 
        Clearly, the {\bf MYTP} is a powerful Gauge Principle which helps 
organize a whole spectrum of phenomena under a single umbrella. For its
applications, only a few examples have been indicated, but its potential 
warrants many more. More importantly, the interlinked 3D-4D structure of 
BS dynamics under {\bf MYTP} [26] premises, gives it access to a whole range 
of physical phenomena, from  spectroscopy to diverse types of loop integrals. 
The emphasis on the spectroscopy sector as an integral part of quark 
physics was first given by Feynman et al [39].

\end{document}